\documentclass[12pt]{article}
\usepackage{epsfig,epsf,amssymb}	%

\usepackage{color}

\begin{document}

\begin{Huge}
\begin{center}
{\bf Unification, the Big Bang and the Cosmological Constant}
\end{center}
\end{Huge}

\begin{large}
\begin{center}
{\bf Edward Tetteh-Lartey}
\linebreak { Dept of Physics, Texas $A\&M$ University, College Station, Texas 77845, USA. Email: lartey@fnal.gov}
\linebreak {\today}
 \end{center}
\end{large}

\vspace{2.0in}

\begin{abstract}
The two major goals in fundamental physics are: 1) Unification of all forces incorporating relativity and quantum theory, 2) Understanding the origin and evolution of the Universe as well as explaining the smallness of the cosmological constant.

Several efforts have been made in the last few decades towards achieving these goals with some successes and failures. 
The current best theory we have for unification of all forces is Superstring/M Theory. However current evidence suggests our Universe is flat and accelerating. A Universe with a positive cosmological constant will have serious implications for string theory since the S-Matrix cannot be well defined and Superstring/M Theory is only formulated in flat Minkowski background. Holographic principle provides a way out as shown by the AdS/CFT and dS/CFT correspondence, but it remains to be proved if it is valid for our non-conformal, non-supersymmetric Universe. Aside from the issue of defining M-Theory in a de Sitter background, why the cosmological constant is so small remains puzzling and needs to be understood. The  ``cosmological constant problem'' has brought physics to a standstill towards any major development and remains currently the most disturbing issue.

Conventional big bang cosmology has not yet produced a satisfactory explanation of the small value of the cosmological constant. An attempt by SuperString/M Theory in this direction is given by the Ekpyrotic/Cyclic model.

The aim of this review is not to introduce any new concepts not already known, but give an overview of current state of affairs in high energy physics, highlighting some successes and failures and making some few suggestions on areas to focus to resolve some of these outstanding issues.
\end{abstract}
 
\clearpage

\section{\bf Introduction}

 Four fundamental forces exist in nature: The electromagnetic force, weak force,
strong force and gravitational force. The search for a unified description of different phenomena has been historically a guiding principle for the progress of physics. It therefore seems natural to search for a theory unifying all four fundamental interactions. A more solid argument for the unification of fundamental forces is provided by the fact that the coupling constants of all four interactions seem to converge at the grand unification scale of about $10^{16}GeV$.

The standard model provides a quantum version of three of the forces and unifies the electromagnetic and the weak force into electroweak theory. But there is no real deep unification of the electroweak and QCD (strong force). The force which is gravity is Einsteins classical theory of relativity. Quantization of gravity has been very difficult due to the nonlinear mathematics on which Einstein based his theory. The nonlinear mathematics of general relativity clashes with the requirement of quantum theory. A necessary condition for renormalizability is absence of negative dimensional coupling constant, but the dimensional coupling of gravity is negative, hence it is nonrenormalisable. The coupling constant grows stronger with energy and ultraviolet divergences appear when we go to arbitrarily high energies and perturbative theory breaks down.    

But is there a need for developing a quantum theory of gravity?. The answer is yes and the reasons are:

\begin{itemize}

\item Quantum Cosmology:

 At low energies and large distances gravity is weak compared to the other forces but at Planck scale energies ($\thicksim 10^{19}GeV$) and Planck scale distances ($\thicksim 10^{-33}cm$), which was the state of our Universe at the big bang, gravity is strong and comparable to the strength of the other forces. Thus quantum mechanics comes into play and a quantum description is necessary. 

\item Singularity Theorems: 

In general relativity, singularities are unavoidable as proved by Hawking and Penrose. Thus general relativity predicts its own breakdown. Experimental evidence from cosmic microwave background radiation indicates that the Universe started from an initial singularity. This singularity cannot be explained by Einstein's classical theory of general relativity hence there is a need for a quantum version.

\item Black Hole Information Paradox:

Hawking's discovery of black hole radiation indicates a contradiction of quantum mechanics and general relativity. General relativity suggest that not even light can get out of a black hole horizon whilst quantum mechanics says we need the information inside the black hole. Thus there is information loss and violation of unitarity. An indication of a possible non-locality in physics. A full quantum theory must explain this puzzle. 

\item Unification of all interactions:

At present all non-gravitational interactions have been accommodated into a quantum framework as presented by the Standard Model. Since gravity couples to all forms of energy as stated by the equivalence principle, it is expected that in a unified theory of all interactions, gravity must be quantized. 

\item Ultra Violet Divergences: 

Quantum field theory is plaque with divergences. It is believed that at small distances (high momenta) space-time is quantized and this divergences can be avoided

\end{itemize}

A correct quantum theory of gravity must, as listed in$~\cite{smo}$:

\begin{itemize}
\item Tell us whether the principles of general relativity and quantum mechanics are true as they stand, or are in need of modification.
\item Give a precise description of nature at all scales, including the Planck scale. 
\item Tell us what time and space are, in a language fully compatible with both quantum theory and the fact that the geometry of space-time is dynamical. Tell us how light cones, causal structure, the metric, etc are to be described quantum mechanically, and at the Planck scale.
\item Give a derivation of the black hole entropy and temperature. Explain how the black hole entropy can be understood as a statistical entropy, gotten by coarse graining the quantum description.
\item Be compatible with the apparently observed positive, but small, value of the cosmological constant. Explain the entropy of the cosmological horizon.
\item Explain what happens at singularities of classical general relativity.
\item Be fully background independent. This means that no classical fields, or solutions to the classical field equations appear in the theory in any capacity, except as approximations to quantum states and histories.
\item Predict new physical phenomenon, at least some of which are testable in current or near future experiments.
\item Explain how classical general relativity emerges in an appropriate low energy limit from the physics of the Planck scale.
\item Predict whether the observed global Lorentz invariance of flat space-time is realized exactly in nature, up to infinite boost parameter, or whether there are modifications of the realization of Lorentz invariance for Planck scale energy and momenta.
\item Provide precise predictions for the scattering of gravitons, with each other and with other quanta, to all orders of perturbative expansion around the semi-classical approximation.
\end{itemize}

Some of the main approaches toward the goal of developing a quantum theory of gravity are$~\cite{kiefer}$

\begin{itemize}
\item Quantum General Relativity:

Here quantization rules are applied to classical relativity. Some examples are

\begin{itemize}
\item Covariant Quantization: Examples include 1) perturbation quantum gravity, where the metric $g_{\mu\nu}$ is decomposed into background part $\eta_{\mu\nu}$ and a small perturbative part, $h_{\mu\nu}$, 2) Effective field theories and renormalization-group approaches, 3) Path integral methods, 4) Dynamic triangulation.
\item Canonical Quantization: Examples include quantum geometrodynamics and loop quantum gravity.
\end{itemize}
\item String Theory: Here fundamental particles are described as one-dimensional objects (see later sections in this review).

\item Quantization of Topology or the Theory of Casual Sets. 
\end{itemize}

The two leading theories in the search for a quantum gravitational theory are loop quantum gravity and superstring theory.

Superstring theory appears to have an edge over loop quantum gravity as the current best quantum gravity theory. In String theory gravity arises naturally as a field with quanta called the graviton. In addition to providing a quantum theory of gravity it also unifies all four interactions as opposed to loop quantum gravity which only provides a quantum gravity theory in four space-time dimensions. String theory has made significant progress on some of the listed requirements above, such as statistical prediction of black hole entropy, and has many attractive features such as prediction of supersymmetry, gravity, and extra dimensions. Also it has no adjustable dimensionless parameters$\footnote{Parameters such as the string coupling are determined by the vacuum expectation value of moduli fields}$. It also unifies of all known interactions with absence of ultraviolet divergences which makes it more predictive than conventional quantum field theory.
 
For Superstring/M Theory to gain full acceptance as a theory giving a unified description of nature and the fundamental forces, it must make testable predictions. Unfortunately the String scale is $\thicksim$ the Planck scale and not accessible to current accelerators. However if supersymmetry breaking occurs at $\thicksim 1 TeV$ as suggested, then supersymmetry could be discovered at the LHC at CERN. Discovering supersymmetry would give string theory a big boost showing that gravity, gauge theory, and supersymmetry that arise from string theory in roughly the same way are all part of the description of nature.

Recent development in cosmology indicate that it will be possible to use astrophysics to perform tests of fundamental theory inaccessible to particle accelerators, namely the physics of the vacuum and cosmic evolution. 

The current best theory we have for explaining the evolution of our Universe is Inflationary/Big Bang theory. This has however not provided a satisfactory explanation of the smallness of the cosmological constant, one of the greatest challenges in physics today. 

Superstring/M Theory has made some headway describing this cosmic evolution and  proposing a solution to the cosmological constant problem by providing the Ekrypotic/Cyclic model. This postulates that our Universe resulted from collision of branes embedded in higher dimensions. There are several evolutionary cycles with a collision occurring in each cycle. The cycle begins with a big bang and ends with a big crunch. It explains that the smallness of the cosmological constant as resulting from the dynamic relaxation of the potential of a quintessence scalar field whose field value also determines the interbrane separation. The potential of this scalar field includes all quantum fluctuations and this total value of potential reduces for every cycle of evolution.
However it has received some criticism from some authors that it is not an alternative to inflationary theory but just another inflationary theory$~\cite{linde}$. 

Increasing evidences from cosmology experiments have shown that our Universe is evolving towards a pure de Sitter space-time. This poses a serious problem for Superstring/M Theory since an S-Matrix cannot be well defined for a de Sitter space-time. 

In this review I will give a brief overview of the concepts behind string theory (see $~\cite{pol}$ for detailed review) and then discuss 
issues related to the cosmological constant. I will then conclude with a suggestion on areas that may lead to a solution of these major problems we are encountering in physics. 

String theory and Cosmology are increasingly becoming broad areas of research
and it is impossible to discuss everything into details in one review. Some topics may not be discussed into details since the main goal of this work is to bring important ideas and developments into one review manual to be easily accessible to the reader thus acting as a good reference source. It is expected that the reader will consult the references for more detailed discussion. 

\clearpage

\section{Superstring/M Theory}

Superstring theory is based on the idea that fundamental objects are not point particles as in particle theories but one dimensional objects called strings. The different vibration modes of the fundamental string which could be either open (have Neumann or Dirichlet boundary conditions) or closed (no boundary conditions), are what we call particles, and which from far look like point particles.
Anomaly cancellations leaves us with only five consistent superstring theories, each living in ten space-time dimensions. These five theories are called Type I, Type IIA, Type IIB, SO(32) heterotic, and $E_{8}\times E_{8}$ heterotic. Type one contains only open strings whilst the others contain both open and closed strings (see Table $~\ref{tab-1}$).

\begin{flushleft}
\begin{table}[htbp]
\begin{small}
\caption{The 5 types of string theory}
\label{tab-1}
\begin{tabular}{|l|c|c|c|c|c|}\hline
                     &Type IIB  &Type IIa &$E_{8}\times E_{8}$ Heterotic &S0(32) Heterotic &Type I SO(32)  \\ \\ \hline
String Type          &closed   &closed &closed &closed &open $\&$closed \\ \\ \hline
No of supercharges     &N=2 (chiral)   &N=2 (non chiral) &N=1 &N=1 &N=1 \\ \\ \hline
10d Gauge Group    &none &none &$E_{8}\times E_{8}$ &S0(32) &S0(32) \\ \\ \hline

D-branes        &-1,1,3,5,7 &0,2,4,6,8 &none &none &1,5,9 \\ \\ \hline
\end{tabular}
\end{small}
\end{table}
\end{flushleft}

In open superstring theories the Hilbert space breaks into two sectors: a Ramond(R) sector with periodic wave functions and Ramond boundary conditions and a Neveu-Schwartz(NS)sector with anti-periodic Neveu-Schwartz boundary conditions. 
After GSO projection all space-time bosonic states arise from the NS sector e.g the eight massless photon states arising from a Maxwell gauge field, and all the space-time fermionic states arise from the R sector. 
The ground states in closed superstrings theories are constructed by tensor products of eight left-moving coordinates of the (NS or R) sector with a eight right-moving coordinates of (NS or R) sector. leading to four sectors NS-NS and R-R which are bosonic states and NS-R, R-NS which are fermionic states. NS-NS massless fields comprises of the graviton $g_{\mu\nu}$, the Kalb-Ramond field $B_{uv}$ and the dilaton $\Phi$. Together we have 128 massless bosonic states and 128 massless fermionic states as required by supersymmetry. All massive states are a result of excitation of these ground states.

The NS-NS bosons of type IIA and type IIB theories are the same but the R-R bosons are different. In type IIA theory the massless R-R bosons include the Maxwell field $A_{\mu}$ and a three-index antisymmetric gauge field $A_{\mu\nu\rho}$. In type IIB theory the massless R-R bosons include a scalar field A, a Kalb-Ramond field $A_{\mu\nu}$, and a totally antisymmetric field $A_{\mu\nu\rho\sigma}$.

Heterotic superstrings are constructed from tensor products of 10 left-moving coordinates of open bosonic string living in 26 space-time dimension, with 10 right moving coordinates of open superstring living in ten dimensions. These results in a theory living ten dimensional space-time.  Enforcing absence of gravitational and gauge anomalies produces the gauge groups SO(32) and $E_{8}\times E_{8}$.   

 By defining strings as one dimensional objects worldlines of particle become world sheets such that space-time is smeared out and there are no singular interaction points thus ultraviolet divergences found in field theories are avoided.
To avoid tachyons (negative mass states) in the spectrum, string theory requires supersymmetry, a space-time symmetry relating fermions and bosons. And for consistency a ten dimensional space-time. 

In string theory one replaces Feynman diagrams with stringy ones and space-time is not really needed. One just needs two-dimensional field theory describing the propagation of strings. Whereas in ordinary physics one talks about space-time and classical fields it may contain, in string theory one talks about an auxiliary two-dimensional field theory that encodes the information. A space-time that obeys its classical equations corresponds to a two-dimensional field theory that is conformally invariant\footnote{That is invariant under changes in how one measures distances along the string}$~\cite{3}$.

In the effective description of string theory, i.e at scales below the Planck scale the massive string states are integrated out since they do not propagate. Thus we are left with only massless modes. The bosonic part of the effective action is the given by$~\cite{4}$:

\begin{equation}
S = S_{univ}+ S_{model} \label{1}
\end{equation}

where $S_{univ}$ does not depend on which of the superstring theories we are are looking at, and $S_{model}$ is model dependent which for Type II strings is given by:

\begin{equation}
S^{II}_{model} = \frac{-1}{2\kappa^{2}}\int d^{10}x\Sigma_{p}\frac{1}{2(p+2)!}
F^{2}_{p+2}  \label{2}
\end{equation}

where $F_{p+2}$ is the field strength of a p+1 form RR gauge field and p is the spatial dimension of an extended object called p-brane that couples electrically to the p+2 form gauge field.

The bosonic fields in the universal sector comprises of the metric, the dilaton and the B field, all massless modes. The action is given by:
 
\begin{equation}
S_{univ} = \frac{1}{\kappa^{2}}\int d^{10}x\sqrt{-G}e^{-2\Phi}\left(R+4(\delta\Phi)_{2}-\frac{1}{12}H^{2} \right)    \label{3}
\end{equation}

$\kappa^{2}\thicksim(\alpha\prime)^{4}$ is the ten dimensional gravitational constant, where $\alpha\prime$ is the slope parameter defining the string tension. 
G is the determinant of the metric $g_{\mu\nu}$, R is the scalar curvature, H is the field strength of the B field $B_{\mu\nu}$, and $\Phi$ is the scalar field called the dilaton.

\subsection{Compactification, Dualities and D-branes}

Our world is four dimensional with broken supersymmetry thus the only way for superstring theory to provide a realistic theory making contact with our world is by compactification of the extra six dimensions and breaking down of supersymmetry. 

There are several methods of compatification: toroidal, orbifolds and orientifolds. The simplest case is toroidal (compact space is a torus) compactification which is the same as Kaluza-Klein compactification in field theory which attempts to unify gauge interactions and gravity.  Curling up a spatial dimension into a circle leads to the spectrum of closed strings having two components: their momentum along the circle is quantized in the form $n/R$, where n is an integer, and winding states $w\alpha^{,}/R$ due to wrapping of the string around the circle m times, where R is radius of compactification, and w is called the winding number. 

The presence of extra-dimensions would not be detected directly if the size of the compact space is of order $10^{-33}cm$\footnote{Just too small to be resolved by the most powerful microscope but could be detected by gravitational effects.}. 
After compactification of the six dimensions, our space-time becomes $M_{4}\times M$, where $M_{4}$ is our four dimensional Minkowski space-time and M is the compact space. Each point in the non-compact space then becomes associated with a tiny ball of six-dimensional space.
 
Compactification on $T^{6}$ preserves too much supersymmetry, but we speculate some minimal supersymmmetry to exist in our 4 dimensional world at energy scales above 1 TeV.
Most choices of $M$ will not yield a consistent string theory since the associated two-dimensional field theory which is now most appropriately described as a non-linear sigma model with target space $M_{4}\times M$- will not be conformally invariant.  

To preserve the minimal amount of supersymmetry N=1 in 4 dimensions, we need to compactify on a special kind of 6-manifold called a Calabi-Yau Manifold. 
A Calabi-Yau Manifold is a complex manifold which admits a metric $g_{\mu\nu}$ whose Ricci tensor $R_{\mu\nu}$ vanishes.
The problem is that there are many Calabi-Yau manifolds each with different physics on $M_{4}$ and not knowing which is the right one to choose leads to lost of predictive power.

Compactification leads to larger degrees of freedom and to several new stringy phenomena such as: winding states, enhanced gauge symmetries, dualities and D-branes. 
When the five consistent string theories are compactified on an appropriate manifold from their ten dimensional space-time, several different string theories emerge in lower dimensions. Each of these theories is parameterized by a set of parameters known as moduli$\footnote{In string theory these moduli are related to vacuum expectation values of various dynamical fields and are expected to take definite values when supersymmetry is broken}$.

\begin{itemize}
\item String coupling constant $g_{s} \sim e^{\Phi}$ (A relation to the vacuum expectation value of the dilaton field $\Phi$)
\item Shape and size of compact manifold $M$ 
\item Various other background fields. 
\end{itemize}

Inside the moduli space of the theory there is a certain region where the string coupling is weak and perturbation theory is valid. Elsewhere the theory is strongly coupled and thus nonperturbative.

\begin{center}
\includegraphics[width=7.5in]{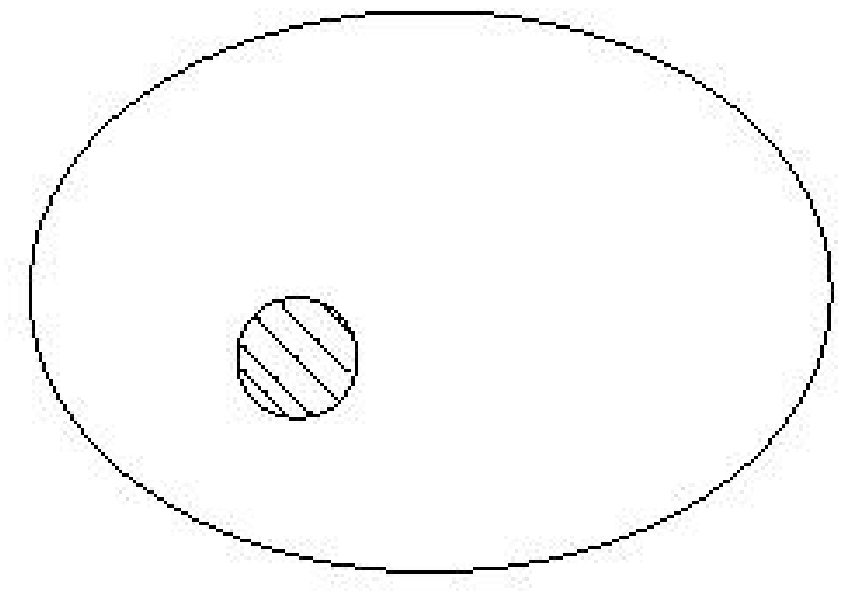} 
{\it Fig. 1. Moduli space of a string theory showing a weak coupling region (shaded region) and a strong coupling region (the white region)}
\end{center} 

\begin{center}
\includegraphics[width=7.5in]{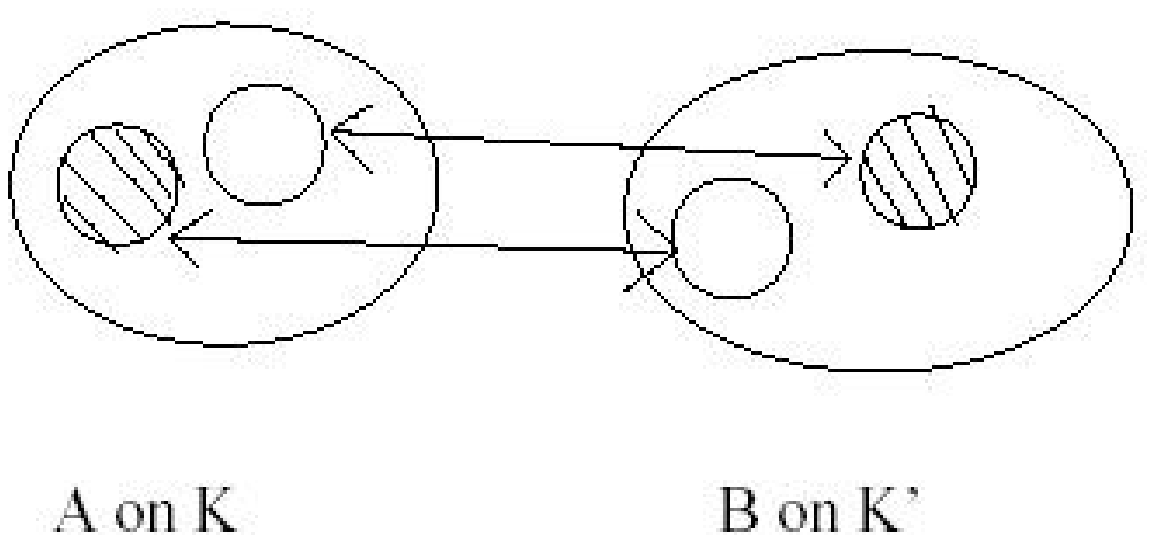} 
{\it Fig. 2. Duality map between the moduli spaces of two different string theories, A on K and B on K' where A and B are two of the five string theories in 10 dimensions, and K, K' are two compact manifolds. This duality gives a mapping between a weak coupling region of one theory (the shaded region) and a strong coupling region of a second theory and vice versa. This is an example of String-string duality}  
\end{center}

Examples of duality symmetries are: T-duality and S-duality, and String-string duality. I discuss each in turn 

\subsubsection{T-duality}

T-duality (or target space duality) transformation maps the weak coupling region of one theory to the weak coupling region of another theory or the same theory (see Figure 3.). 

\begin{center}
\includegraphics[width=7.5in]{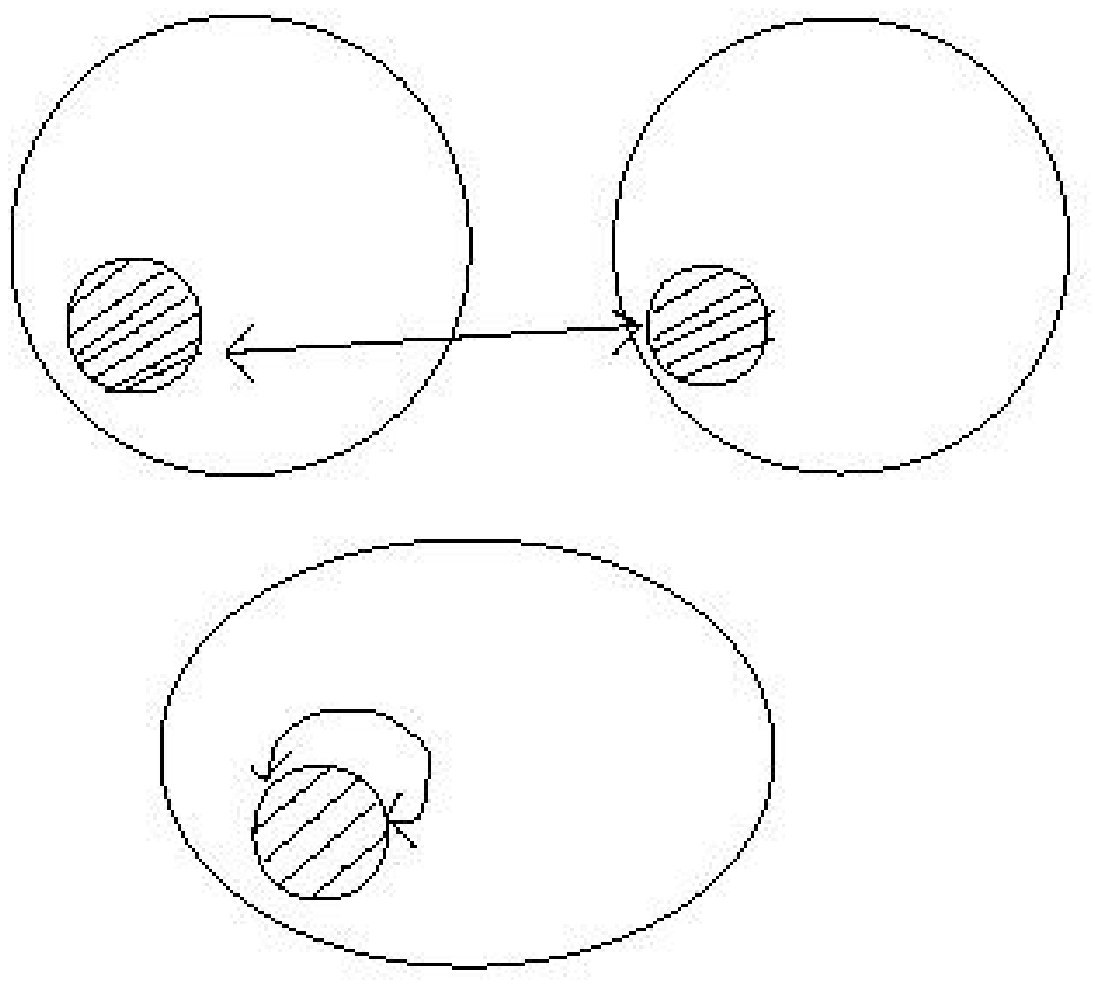} 
{\it Fig. 3. Examples of T-duality relating a weakly coupled theory to a different or the same weakly coupled theory}
\end{center} 

Examples of T-duality are:

Het on $S^{1}$ with radius $\frac{R}{\alpha\prime}$ \ $\stackrel{T-dual}{\leftrightarrow}$ \ Het on $S^{1}$ with radius $\frac{\sqrt{\alpha\prime}}{R}$

IIA on $S^{1}$ with radius $\frac{R}{\alpha\prime}$ \ $\stackrel{T-dual}{\leftrightarrow}$ \ IIB on $S^{1}$ with radius $\frac{\sqrt{\alpha\prime}}{R}$

Heterotic string theory compactified on a circle of radius R is dual to the same theory compactified on radius $R^{-1}$ with same coupling constant.  
Type IIA string theory compactified on a circle of radius R is dual to IIB string theory compactified on a circle of radius $R^{-1}$ at the same value of the coupling constant.
 
The physics when the circle has radius R is indistinguishable from the physics of a circle of radius  $\alpha^{,}/R$. As $R\rightarrow \infty $, winding states become infinitely massive, while compact momenta goes to a continuous spectrum as with the non-compact dimensions. For the case of $R\rightarrow 0 $, the states with compact momentum become infinitely massive but the spectrum of winding states now approaches a continuum\footnote{It does not cost much energy to wrap a string around a small circle}.

Thus as the radius goes to zero the spectrum again seems to approach that of a non-compact dimension. This implies limits $R\rightarrow 0$ and $R\rightarrow \infty$ are physically equivalent.
For open strings, in the limit $R\rightarrow 0$, strings with Neumann boundary conditions have no comparable quantum number to w. Thus states with nonzero momentum go to infinite mass, but no new continuum of states. 

\subsubsection{S-duality}

S-duality or self duality gives an equivalence relation different regions of the moduli space of the same theory. It maps the weak coupling regime of one string theory to the strong coupling region of the same theory, see Figure 4.

This duality was first conjectured in the context of compactification of heterotic string to four dimensions$~\cite{sdua}$. It has also been found that Type II B superstring in ten dimensions is S-dual to itself$~\cite{sdua2}$.

\begin{center}
\includegraphics[width=7.5in]{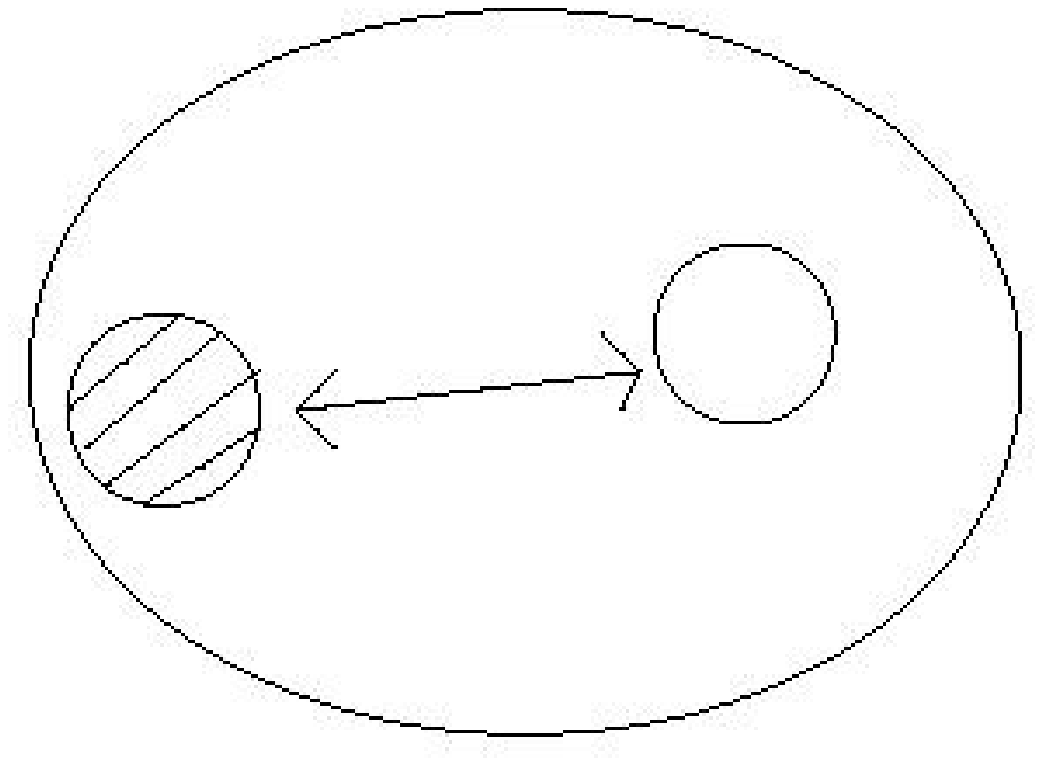} 
{\it Fig. 4. A representation of the moduli space of a self-dual theory. The weak and strong coupling regions of the same theory are related by duality.} 
\end{center} 

For both the heterotic and Type IIB string theories the transformations acts via an element of SL(2,Z) on a complex scalar $\lambda$, whose imaginary part's VEV is related to the coupling constant of the string theory ($g_{s}\sim e^{<\Phi>}$)$~\cite{lust}$.

\subsubsection{String-string duality}

This is a duality relation between different string theories in a way that the perturbative regime of one theory is equivalent to to the non-perturbative regime of the other i.e we have a mapping between the elementary excitations of one theory to the solitonic excitations of the other theory and vice versa. Examples are$~\cite{lust}$:

Het on $T^{4}$ \ $\leftrightarrow$ \ IIA on k3 

Het with gauge group SO(32) in d=10 \ $\leftrightarrow$ \ I in d=10

 In general the duality can relate not just two theories but a whole chain of theories as illustrated in Figure 5.

\begin{center}
\includegraphics[width=7.5in]{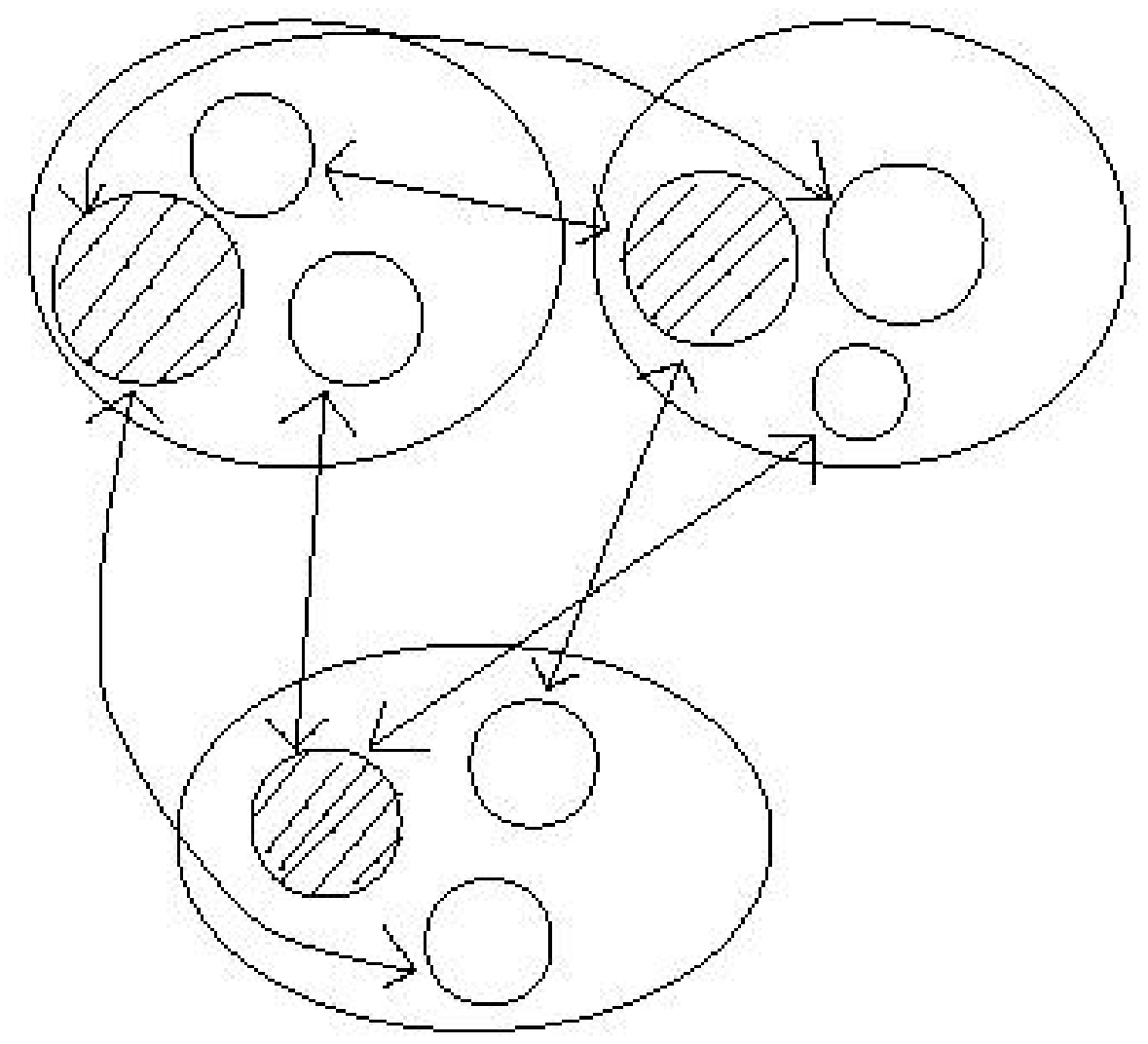}
{\it Fig. 5.  The moduli spaces of a chain of theories related by duality in diagrammatic representation. In each case the shaded region denotes the weak coupling region.} 
\end{center}

\subsection{D-branes}

When an open string theory is compactified on a small torus, the physics is described by a compactification on a large torus but with the  open string endpoints restricted to lie on subspaces. These subspaces or hyperplanes are dynamical objects called Dirichlet membrane or D-brane. Thus by taking compactification and taking the limits $R\rightarrow 0$ and T-duality transformations we have interchanged Neumann boundary conditions with Dirichlet boundary conditions resulting in appearance of nonperturbative objects called Dp-branes in a non-compact space, where p is number of its space dimensions$~\cite{pol}$. Strings carry electric ``charge'' by coupling directly to NS-NS tensor $B_{\mu\nu}$, but couple only to the field strengths of R-R fields and not their potentials, thus are R-R neutral. However Dp-branes couples to R-R fields and thus complements string theory with nonperturbative states\footnote{For nonperturbative objects such as D-branes and their magnetic dual, NS Branes, the interaction amplitude varies inversely to the string coupling and given by $A\sim e^{-1/g_{s}}$, $A\sim e^{-1/g_{s}^{2}}$ respectively. The masses also vary in the same way. Thus as we go to strong coupling, they become light objects. This inverse variation of the amplitude with string coupling allows us to have meaningful values of their interaction probabilities.} carrying R-R charges. 
 
D-branes are BPS states i.e belong to the short multiplet representation. In supersymmetric theories, BPS states are stable objects i.e are invariant under a subset of supersymmetry transformations as the string coupling changes from small to large, i.e moving from perturbative to non-perturbative region. Dp with p=-1 are called D instanton, D0-brane is a zero dimensional object, D-string is a one dimensional brane, D2-brane is called a membrane. The standard models particles are confined on D-branes. Dp-branes are p-branes with Dirichlet boundary conditions on which open strings can end.

Dp-branes couple to p+1 dimensional objects which gives the charge. D-branes are hodge dual to NS five branes which are their magnetic dual and have magnetic charges. Type IIA theories have RR gauge fields $A_{\mu}$, $A_{\mu\nu\rho}$ and $A_{\mu\nu\rho\sigma\lambda}$ and hence couple to D0-branes, D2-branes, D4-branes etc. Type IIB has R-R gauge fields A, $A_{\mu\nu}$,  and $A_{\mu\nu\rho\sigma}$ so they couple to D(-1)-branes, D1-branes, D3-branes. Tension on a D-brane is given by $T\thicksim 1/g_{s}$, where $g_{s}$ is the string coupling. Thus in the strong coupling limit D-branes become light objects. D-branes are dynamic objects on which open strings live on. The action for D-brane is given in Appendix A.

   We gain insights into non-perturbative effects in string theory by finding the BPS states of perturbative string theory. This shows the usefulness of D-branes.

Discovery of D-branes has made a remarkable impact in String theory, some of which are:

\begin{itemize}

\item Discovery of nonperturbative string dualities. 
\item A microscopic explanation of black hole entropy and the rate of emission of thermal (Hawking) radiation for black holes in string theory.
\item AdS/CFT correspondence. A holographic principle conjectured by Maldacena. Will be discussed in later sections.
\item Probes of short-distances in space-time, where quantum gravitational fluctuations become important and classical general relativity breaks down.
\item Modeling our world as a D-brane. This may be used to explain why gravity couples so weakly to matter, i.e. why the effective Planck mass in our (3+1) dimensional world is so large, and hence gives a potential explanation of the hierarchy problem $m_{p}\gg m_{weak}$. 
\item Ekpyrotic/Cyclic model of the Universe: Here branes are used to explain the cosmic evolution of the Universe as a proposed alternative to inflational theory. Also explains the source of the small cosmological constant.
\end{itemize}

\subsection{M-theory}

The five consistent string theories were thought to be far to many for a theory that is suppose to be unique and unifying all forces.
A second string revolution started around 1995 with the discovery of duality symmetries.  This allowed string theories to be extended beyond perturbative expansion to probe nonperturbative features. The three major implications of these discoveries were$~\cite{zabo}$:
 
\begin{itemize}
\item  Dualities relate all five superstring theories in ten dimensions to one another.

The different theories are just perturbative expansions of a unique underlying theory ${\it  U}$ about five different, consistent quantum vacua. The implication is that there is a complete unique theory of nature, whose equation of motion admits many vacua.

\item The theory ${\it  U}$ also has a solution called ``M-Theory'' which lives in 11 space-time dimensions

The low-energy limit of M-Theory is 11-dimensional supergravity. All five superstring theories can be thought of as originating from M-Theory (see Figure. 6). The underlying theory ${\it  U}$ is shown in Figure 7.

\item In addition to the fundamental strings, the theory ${\it U}$ admits a variety of extended nonperturtabive excitations called ``p-branes''.

\end{itemize}

\begin{center}
\includegraphics[width=7.5in]{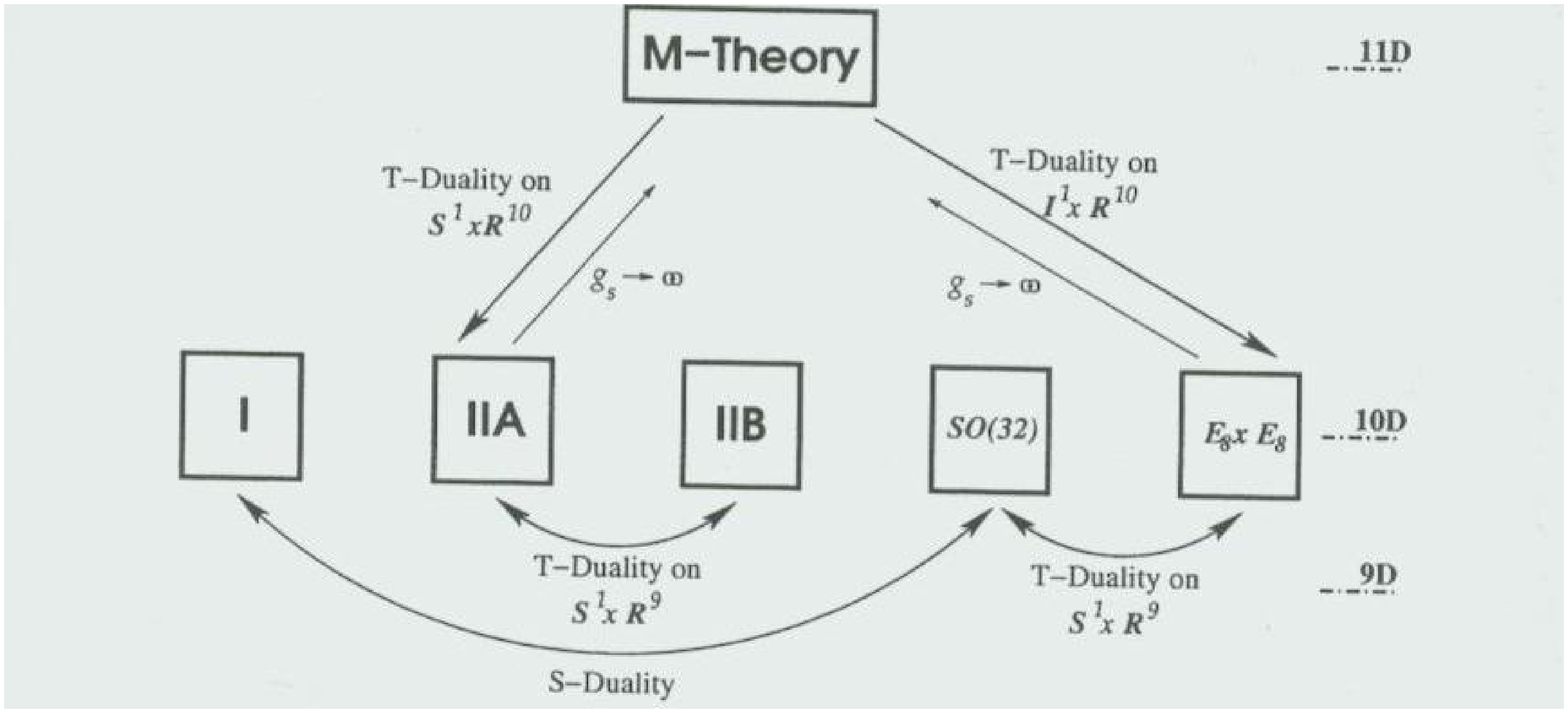} 
{\it Fig. 6. The various duality transformations that relate the supertstring theories in nine and ten dimensions. T-Duality inverts the radius R of the circle $S^{1}$, or the length of the finite interval $I^{1}$, along which a single direction of the spacetime is compactified, i.e. $R\rightarrow l^{2}_{p}/R$. S-duality inverts the (dimensionless) string coupling constant $g_{s}$, $g_{s}\rightarrow 1/g_{s}$, and is the analog of electric-magnetic duality (or strong-weak coupling duality) in four dimensional gauge theories. M-Theory originates as the strong coupling of either the Type IIA or $E_{8}\times E_{8}$ heterotic string theories. }
\end{center} 

\begin{center}
\includegraphics[width=7.5in]{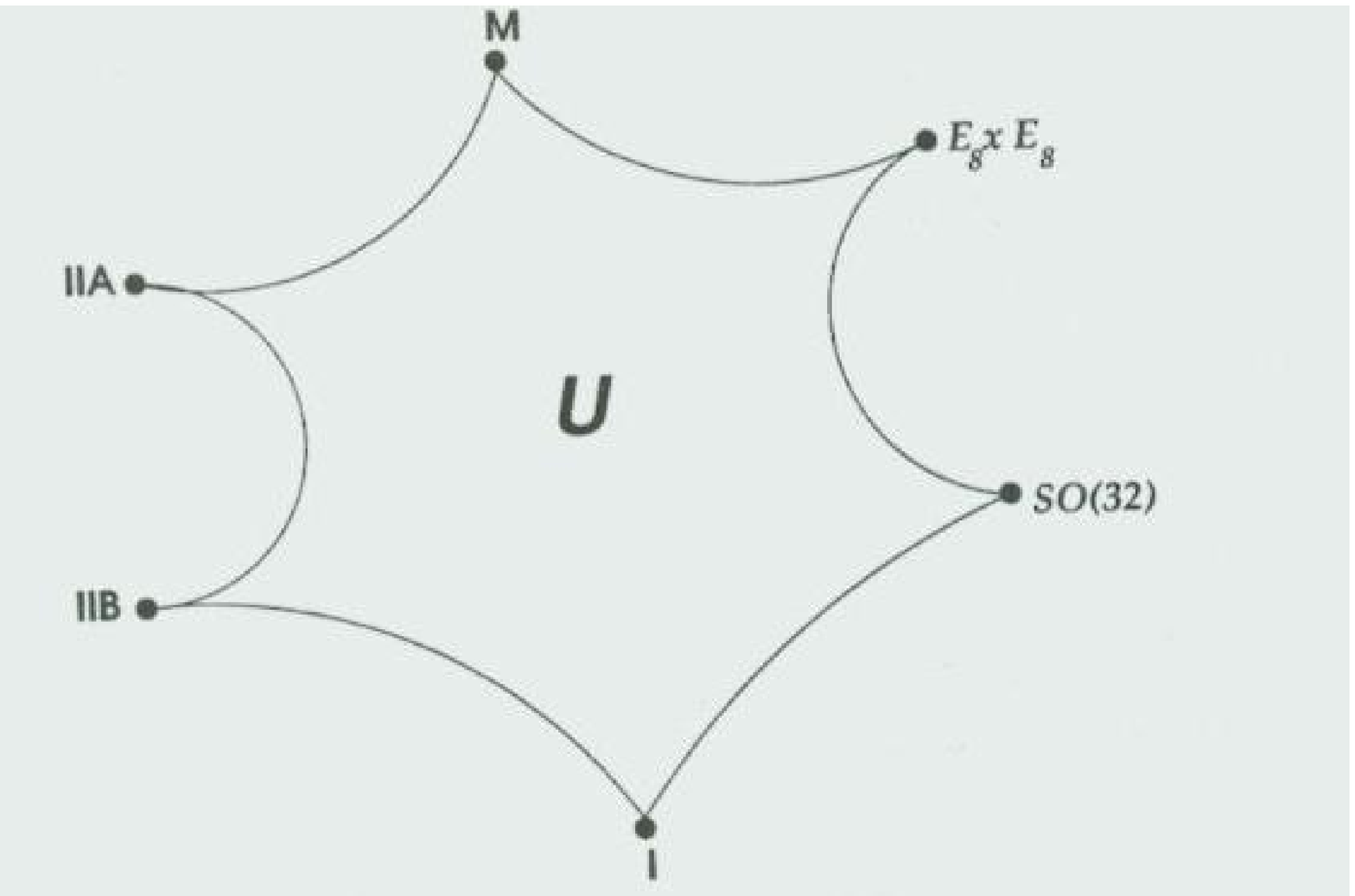} 
{\it Fig. 7.  The space U of quantum string vacua. At each node a weakly-coupled string description is possible}
\end{center}

In the low energy limit, the various superstring theories are described by supergravity theories. The low-energy effective theory of Type IIA string theory is ten dimensional Type IIA supergravity and the low energy limit of Type IIB string theory is Type IIB supergravity. D-branes can be found as solutions to Type II string and massive supergravity theories$~\cite{4}$. 
 Type IIA supergravity can also be obtained by dimensional reduction of supergravity in eleven dimensions.

The eleven dimensional supergravity multiplet contains the following massless fields: a metric $G_{MN}$, a three-form potential $A_{3}$ with components $A_{MNP}$ and a Majorana gravitino $\Psi_{M}$.  It has a total of 256 degrees of freedom (dofs); 128 bosonic and 128 fermionic. 

The bosonic part of the eleven dimensional supergravity action is given by$~\cite{adel}$

\begin{equation}
S_{bos}^{11}=\frac{1}{2}\int d^{11}x \sqrt{G}(R + |dA_{3}|^{2})
+ \int A_{3}\wedge dA_{3} \wedge dA_{3} \label{sup}
\end{equation}

with the fermionic terms determined by supersymmetry.

Reducing this theory to ten dimensions by compactifying the eleven dimension $x^{11}$, on a circle, the eleven dimensional Majorana gravitino (a supersymmetric pair of the graviton) 

$\Psi \equiv \left( \begin{array}{c}  \psi^{1}_{M} \\ 
\psi^{2}_{M} \end{array} \right) $ 

gives rise in ten dimensions to a pair of Majorana-Weyl gravitinos (of opposite chiralty) $\psi^{a}_{\mu}$,
and a pair of Majorana-Weyl spinors $\psi^{a}\equiv \psi^{a}_{11}$ a=1,2. 
The eleven dimensional three form gives rise in ten dimensions to a three form $A_{\mu\nu\rho}$ (56 dofs) and a two-form $B_{\mu\nu} \equiv A_{\mu\nu 11}$ (28 dofs), while the eleven dimensional metric gives in ten dimensions a metric $G_{\mu\nu}$ (35 dofs), a scalar $e^{2\gamma}\equiv G_{11 11}$, and a vector potential $A_{\mu} \equiv -e^{-2\gamma}G_{\mu 11}$ (8 dofs), again a total of 128 bosonic dofs.  
The eleven dimensional metric is given by:

\begin{equation}
ds^{2} = G_{MN}dx^{M}dx^{N} \\
       = G_{\mu\nu}dx^{\mu}dx^{\nu} + e^{2\gamma}(dx^{11} - A_{\mu}dx^{\mu})^{2}, \label{sup2}
\end{equation}

and resulting ten dimensional action after compactification is given by:

\begin{equation}
\int d^{10}x\sqrt{G_{(10)}}\left[e^{\gamma}(R+|\bigtriangledown\gamma|^{2} + |dA_{3}|^{2}) + e^{3\gamma}|dA|^{2} + e^{-\gamma}|dB|^{2}\right] + \int B\wedge dA_{3} \wedge dA_{3} \label{sup4}
\end{equation}

The usual form of this IIA supergravity action is:

\begin{equation}
\int d^{10}x\sqrt{g}\left[e^{-2\phi}(R + |\bigtriangledown\gamma|^{2} + |dB|^{2}) + |dA_{3}|^{2} + |dA|^{2}\right] + \int B\wedge dA_{3} \wedge dA_{3}. \label{sup5}
\end{equation}  

Compactification to ten dimensions not only results massless modes forming a supermultiplet with 256 states but also massive Kaluza-Klein (KK) modes.
Thus each massless mode has a corresponding  Kaluza-Klein massive mode.
With a given compactification radius $R_{11}$, the KK modes have momenta $n/R_{11}^{(g)}$ and are the only massive states in ten dimensional supergravity. 
Their mass is related to the string coupling by:

\begin{equation}
M=\frac{n}{R_{11}^{g}}=\frac{n}{\sqrt{\alpha'}g_{s}}=\frac{n}{l_{s}g_{s}}.\label{sup7} 
\end{equation}

Hence at large coupling become very light objects.
 
All KK states of the full eleven dimensional supergravity on $R^{10}\times S^{1}$ are contained in the Type IIA superstring theory, with each state being actually a full supermultiplet of 256 states, these are the D0-branes. In the strong coupling limit $g_{s}\rightarrow \infty$, Dp-branes become massless ($M=1/l_{s}g_{s}$) and are all low energy states of the IIA superstring theory. Also as  $g_{s}\rightarrow \inf$, the compactification radius $R_{11}\rightarrow \infty$ and one gets uncompactified eleven dimensional supergravity. Thus it is conjectured that eleven dimensional supergravity is the low energy limit of IIA superstring theory at strong coupling  with  $R_{11}\sim g_{s}$. 

Eleven dimensional supergravity is the low-energy limit of some consistent theory, called M-Theory which describes the strong coupling limit of IIA superstring. M-Theory with its eleventh dimension compactified on a circle of radius $R_{11}$ is identical to IIA superstring theory with string coupling $g_{s}=R^{(g)}_{11}/\sqrt{\alpha^{'}}$ where $R_{11}^{g}=g_{s}^{1/3}R_{11}$ is the eleven dimensional radius when measured with the string metric g. 

M-theory describes IIA superstring which have D0, D2, D4, D6 and D8 branes as well as fundamental string (F1 brane). 
Since M-Theory contains a third-rank gauge field $A_{MNP}$ it must have a 2-brane to which the gauge field couples. The dual potential of this gauge field has rank 6, and thus couples to a 5-brane\footnote{A rank (p+1) gauge field couples to a p-brane. It hodge dual is a gauge field with rank-(D-p-3) and thus couples to a D-p-4-brane. D is the full target space-time dimensions in which the brane propagates, p is the space dimension of the brane.}. Thus in M-Theory we expect to have a 2-brane and a 5-brane in addition to the  gravitino, and the three-form potential, M-theory .

For M-theory to make contact with our four dimensional world, 7 extra dimensions would have to be compactified  with preservation of some supersymmetry. One can get 4d N=1 theories from M-Theory by compactifying on a 7-manifold of $G_{2}$ holonomy$~\cite{g2holo}$.
The first example of a Kaluza-Klein compactification with non-trivial holonomy was provided by the squashed $S^{7}$ which has $G^{2}$ holonomy and yielded N=1 in D=4 space-time$~\cite{awa, duff}$.
Examples of such theories have been studied$~\cite{g2st}$ but not much is known about their physics. 

Despite these efforts in String/M-Theory compactifications, no complete quantitative agreement has been found with the various elementary particles of the standard model such as the masses.

Over the past few years development of M-Theory has lead to many useful insights in physics some of which are:

\begin{itemize}
\item Matrix models $~\cite{mat}$
\item AdS/CFT correspondence $~\cite{ads}$
\item Non-commutative geometry $~\cite{ncom}$
\item F-Theory $~\cite{ftheo}$
\item K-Theory $~\cite{ktheo}$
\item $E_{11}$ symmetry $~\cite{esym}$
\item Topological strings and twistors $~\cite{tstr}$
\end{itemize}

Here I give a brief review of the Matrix Model and the AdS/CFT Correspondence.

\subsubsection{Matrix Model}

The matrix model by Banks, Fischler, Shenker and Susskind (BFSS) is a conjecture that states that M-Theory in the light cone frame (infinite momentum frame), is exactly described by the large N Limit of a particular supersymmetric matrix quantum mechanics$~\cite{matmod}$.

As stated in the previous section, M-Theory is the strong coupling limit of Type IIA string theory  The KK states of the eleven dimension supergravity corresponds to bound states of N D0-branes. The D0 branes are point particles which carry a single unit of RR charge and longitudinal momentum $P_{11}=1/R$. D0 branes carry the quantum numbers of the first massive KK modes of the basic eleven dimensional supergravity multiplet, including 44 gravitons, 84 components of a 3-form and 128 gravitinos. Collectively all three types are called supergravitons.

From Appendix A, we know that a collection of N Dp-branes is described by ten dimensional U(N) super Yang-Mills theory reduced to p+1 dimensions, i.e $N \times N$ hermitian matrix quantum mechanics. For p=0, a collection of N D0 branes can be described by dimensional reduction of ten dimensional U(N) super Yang-Mills theory reduced to 0+1 dimensions. The action is given by:

\begin{equation}
S_{D_{0}}= \frac{1}{2g_{s}\sqrt{\alpha\prime}}\int d\tau Tr \left( \dot{\Phi}^{m}\dot{\Phi}_{m} + \frac{1}{(2\pi\alpha\prime)^{2}}\sum_{m<n}[\Phi^{m},\Phi^{n}]^{2} + 
\frac{1}{2\pi\alpha '}\theta^{T} i \dot{\theta} - \frac{1}{(2\pi\alpha\prime)^{2}} \theta^{T}\Gamma_{m}[\Phi^{m},\theta] \right),  \label{dbrane1}
\end{equation} 

What the BFSS theory is then saying is that in the limit $N \rightarrow \infty $ (large collection of Dp-branes with gauge theory U(N)), the above equation becomes M-Theory in the light cone frame. 
To prove this we must first define M-Theory in the Infinite Momentum Frame (IMF) or Light Cone Frame (LCF).

For a collection of particles, IMF is defined to be a reference frame in which the total momentum P, is very large. All individual momenta can be written as

\begin{equation}
p_{a}=\eta_{a}P + p_{\perp}^{a} \label{dbrane2}
\end{equation}

with $p_{\perp}^{a}.P=0$, \ $\sum p_{\perp}^{a}=0$, \ and $\sum \eta_{a}=1$.

implying the observer is moving with high velocity in the -P direction. For sufficiently large boost, all $\eta_{a}$ are strictly positive. 
The energy of any particle is then given by$~\cite{adel}$:

\begin{equation}
E_{a}=\sqrt{p_{a}^{2}+m_{a}^{2}} \\
=\eta_{a}P + \frac{(p_{\perp}^{a})^{2} + m_{a}^{2}}{2\eta_{a}P} + O(P^{-2}) \label{dbrane3}
\end{equation}

This equation has the non-relativistic structure $(p_{\perp}^{a})^{2}/2\mu_{a}$ of a d-2 dimensional system with the role of the non-relativistic masses $\mu_{a}$ played by $\eta_{a}P$, with the only difference being the constant, $\eta_{a} P + \frac{m_{a}^{2}}{2\eta_{a}P}$.

Before proceeding with the prove it is worth noting that infinite momentum frame could also be interpreted as the light cone frame since they are similar. And here is the similarity: In the light cone frame
we single out one spatial direction called longitudinal with momentum $p_{L}^{a}=\eta_{a}P$ and define $p_{\pm}^{a}=E^{a}\pm p_{L}^{a}=E^{a}\pm\eta^{a}P.$ 
Then the mass shell condition reads $p_{-}^{a}p_{+}^{a}-(p_{\perp}^{a})^{2}=m_{a}^{2}$ or

\begin{equation}
E_{a}-\eta_{a}P=\frac{(p_{\perp}^{a})^{a}+m_{a}^{2}}{p_{+}^{a}}  \label{dbrane4}\end{equation}

If P and hence $p_{L}^{a}=\eta_{a}P$ is large, one has $E^{a}\approxeq \eta_{a}P$ and $p_{+}^{a}\approxeq 2\eta_{a}P$ and which agrees with eq.$~\ref{dbrane3}$ taken in that limit.

Considering M-Theory in the IMF, we separate the components of the eleven dimensional momenta as follows: $p_{0},p_{i}, i=1,....9$ and $p_{11}$. 

We then boost in the 11th direction to the IMF until all $p_{11}^{a}$  become positive. 
The eleventh dimension $x^{11}$, is then compactified on a circle of radius R. This results in the quantization of all momenta $p_{11}^{a}$ as $n_{a}/R$ with $n_{a}>0$. Since there are no eleven dimensional masses $m_{a}$ the energy momentum relation becomes:

\begin{equation}
E - p_{11}^{tot} = \sum_{a}\frac{(p_{\perp}^{a})^{2}}{2p_{11}^{a}} \label{dbrane5}\end{equation}

The above equation exhibits the non-relativistic structure we saw in eq$~\ref{dbrane3}$.
At this point we have M-Theory in an infinite momentum frame with full Galilean invariance in the transverse dimensions.

We have pointed out previously that KK modes appear after compactification and are the Dp-branes in IIA superstring. The RR photon that couples to a D0-brane in the IIA superstring is the KK photon that results from compactifying $x^{11}$ on $S^{1}$ of radius R with the RR charge corresponding to $p_{11}$.

A single D0-brane carries one unit of RR charge and thus is given by $p_{11} = 1/R$. 
It fills out a whole supermultiplet of 256 states. Since in eleven dimensions it is massless (graviton multiplet), in ten dimensions it is BPS saturated. There are also KK states with $p_{11}=N/R$, N being an arbitrary integer.
N>1 are bound states of N D0-branes, while $N < 0$ corresponds to anti-DO-branes or bound states\footnote{Note that DO-branes with RR charge 1/R = KK massive modes is compactified M-Theory with RR charge 1/R. The RR are the gauge photons in both cases. In M-Theory they originate from compactifying the metric tensor. Note that the KK massive modes are the massive modes of the graviton and gravitino i.e the metric in 10D}. 
As mentioned earlier, taking the total $p_{11}$ to infinity to reach the IMF limit, leaves only positive $p_{11}$, i.e $N > 0$. This means M-Theory in the IMF should only contain D0-branes and their bound states. The anti-D0-branes get boosted to infinite energy and have somehow implicitly been integrated out.

So now we have arrived at our M-Theory (in IMF) being described by D0-branes quantum mechanics.  
The membranes (i.e 2-branes) and 5-branes in M-Theory can also be described within the D0-brane quantum mechanics$~\cite{adel}$. 

Restating the BFSS conjecture: M-Theory in the IMF is a theory in which the only dynamical degrees of freedom are D0-branes each of which carries a minimal quantum of $p_{11}=1/R$. This action is described by the effective action for N D0-branes which is a particular $N \times N$ matrix quantum mechanics, to be taken in the $N \rightarrow \infty $ limit.

Though M-theory has had some setbacks as a true unified theory of nature one successful area is the accurate counting of microstates (entropy) for certain highly supersymmetric black holes saturating the Bogomolnyi-Prasad-Somerfield (BPS) bound by Stominger and Vafa$~\cite{vf}$.

\subsubsection{AdS/CFT Correspondence}

Anti-deSitter space-time is the maximally symmetric solution of Einstein's equations with a negative cosmological constant $\Lambda < 0$. The metric satisfies the Einstein equations with $\Lambda < 0$. A pure anti-deSitter is one in absence of matter fields i.e the vacuum solution.
It is the the most symmetric space-time with negative curvature. 

In a remarkable development, Maldacena (1997) conjectured that the quantum field theory that lives on a collection of D3-branes (in the IIB theory) is actually equivalent to Type IIB string theory in the geometry that the gravitational field of the D3-branes creates.

The duality in its full form is stated as:

``Four-dimensional {\it N = 4} supersymmetric SU(N{c}) gauge theory is equivalent to IIB string theory with $AdS_{5} \times S_{5}$ boundary conditions.''

Maldacena arrived at this conjecture by considering a stack of $N_{c}$ parallel D3-branes on top of each other. 
Each D3-brane couples to gravity with a strength proportional to the dimensionless string coupling $g_{s}$, so the distortion of the metric by the branes is proportional to $g_{s}N_{c}$. When $g_{s}N_{c}\ll 1$ the spacetime is nearly flat and there are two types of string excitations: 1) open strings on the brane whose low energy modes are described by a $U(N_{c})$ gauge theory and 2) close strings away from the brane. When $g_{s}N_{c}\gg 1$, the back-reaction is important and the metric describes an extremal black 3-brane. Near the horizon the spacetime becomes a product of $S_{5}$ and $AdS_{5}$.\footnote{This is directly analogous to the fact that near the horizon of an extremal Reissner-Nordstrom black hole, the spacetime is $AdS_{2}\times S_{2}$.}

String states near the horizon are strongly red-shifted and have very low energy as seen asymptotically. In a certain low energy limit, one can decouple these strings from the strings in the asymptotically flat region. At weak coupling $g_{s}N_{c}\ll 1$, this same limit decouples the excitations of the 3-branes from the closed strings. Thus the low energy decoupled physics is described by the gauge theory at small $g_{s}$ and by the $AdS_{5}\times S_{5}$ closed string theory at large $g_{s}$. The simplest conjecture is that these are the same theory as seen at different values of the coupling$\footnote{The U(1) factor in $U(N_{c})= SU(N_{c})\times U(1)$ also decouples, it is Belia and does not feel the strong gauge interactions}$.
The fact that very different gauge theory and gravity calculations were found to give the same answers for a variety of string-brane interactions is quite remarkable. 

\section{Cosmic Evolution}

On cosmological scales, a description of our Universe on the symmetries of homogeneity and isotropy,\footnote{Homogeneity implies the same at every point, i.e translational invariant. Isotropy implies the same in every direction, rotational invariant. The implication of the homogeneity and isotropy is what we call the Cosmological Principle} leads to the Robertson-Walker metric as a solution of the Einstein field equation:

\begin{equation}
G_{\mu\nu} = 8\pi GT_{\mu\nu}, \label{cosmoeq1}
\end{equation}

\begin{equation}
R_{\mu\nu} - \frac{1}{2}g_{\mu\nu}R = 8\pi GT_{\mu\nu}. \label{cosmoeq1a}
\end{equation}

where $g_{\mu\nu}$ is the space-time metric and R is the Ricci scalar curvature.
$T_{\mu\nu}$ is a stress energy tensor describing the distribution of mass in space, G is Newton's gravitational constant and the Einstein Tensor $G_{\mu\nu}$ is a complicated function of the metric and its first and second derivatives.
 
The Robertson-Walker metric satisfying this equation is given by: 

\begin{equation}
ds^{2} = -dt^{2} + a^{2}(t) \left[ \frac{dr^{2}}{1 - kr^{2}} + r^{2}(d\theta^{2} + sin^{2}\theta d\phi^{2}) \right],   \label{cosmoeq2}
\end{equation}

where the scale factor $a^{2}(t)$ describes the relative size of spacelike hypersurfaces at different times and contains all the dynamics of the Universe, k is a constant describing the curvature of space: k =0 for flat hypersurfaces (flat Universe), k = -1 for negatively curved hypersurfaces (open Universe), and k = +1 for positively curved hypersurfaces (closed Universe).

If the contents of the Universe is modeled as a perfect fluid with density $\rho$, and pressure p, the stress-energy tensor is given by:

\begin{equation}
T_{\mu\nu} = (\rho + p)U_{\mu}U_{\nu} + pg_{\mu\nu} \label{cosmoeq3a}
\end{equation}

where $U^{u}$ is the four-velocity of the fluid\footnote{To obtain a Robertson-Walker solution to Einstein equation, the rest frame of the fluid must be that of a co-moving observer in the metric given above}.

Inserting the Robertson-Walker metric into Einstein's equations yields the Friedman equations:

\begin{equation}  
\left(\frac{\dot{a}}{a} \right)^{2} = \frac{8\pi G}{3}\rho - \frac{k}{a^{2}}, \label{cosmoeq3}
\end{equation}

and 

\begin{equation}  
\frac{\ddot{a}}{a} = -\frac{4\pi G}{3}(\rho + 3p) \label{cosmoeq4}.
\end{equation}

The equation of state relates the pressure p and the density $\rho$ of a fluid by, $p = \omega\rho$. This is a simple equation of state satisfied by most fluids.
Note that the second derivative of the scale factor depends on the equation of state of the fluid. For $\omega > 0$, the pressure will be positive and $\ddot{a}<0$ (expansion will decelerate), whilst for  $\omega < 0$, $\ddot{a}>0$ and we have accelerated expansion. 

$H =\dot{a}/a$ is the Hubble parameter which characterizes the rate of expansion of the Universe. Its value at the present epoch is the Hubble constant, $H_{0}$.The density parameter in a species i is given by:

\begin{equation}
\Omega_{i} = \frac{8\pi G}{3H^{2}}\rho_{i} = \frac{\rho_{i}}{\rho_{crit}}  \label{cosmoeq5},
\end{equation}

where the critical density is defined by:

\begin{equation}
\rho_{crit} = \frac{3H^{2}}{8\pi G}  \label{cosmoeq6},
\end{equation}

corresponding to the energy density of a flat Universe. In terms of the total density parameter

\begin{equation}
\Omega = \sum_{i}\Omega_{i}, \label{cosmoeq7}
\end{equation}

the Friedman equation$(~\ref{cosmoeq3})$ can be written as:

\begin{equation}
\Omega - 1 = \frac{k}{H^{2}a^{2}}. \label{cosmoeq8}
\end{equation}

The equation of state parameter $\Omega$ determines the sign of k as shown below $~\cite{sca}$. 

\begin{equation}
\begin{array}{ccccccc}
\rho < \rho_{crit} &\leftrightarrow &\Omega < 1 &\leftrightarrow &k = -1 &\leftrightarrow  &open \\
\rho = \rho_{crit} &\leftrightarrow &\Omega = 1 &\leftrightarrow &k = 0 &\leftrightarrow  &flat \\
\rho > \rho_{crit} &\leftrightarrow &\Omega > 1 &\leftrightarrow &k = +1 &\leftrightarrow  &closed 
\end{array}
\end{equation}

Note that $\Omega_{i}/\Omega_{j} = \rho_{i}/\rho_{j} = a^{-(n_{i}-n_{j})},$ so that the relative amounts of energy in different components will change as the Universe evolves.

A photon traveling through an expanding Universe will undergo a redshift of its frequency proportional to the amount of expansion. The redshift z, is used as a way of specifying the scale factor at a given epoch:

\begin{equation}
1+ z = \frac{\lambda_{obs}}{\lambda_{emitted}} = \frac{a_{0}}{a_{emitted}}
\end{equation}

where the subscript 0 refers to the value of a quantity in the present Universe, and $\lambda$ is the wavelength of the photon. 

Einstein's equations relate the dynamics of the scale factor to the energy-momentum tensor. For many cosmological applications we assume that the Universe is dominated by a perfect fluid, in which case the energy-momentum tensor is specified by an energy density $\rho$ and pressure p: $T_{00} = \rho$,  \  $T_{\mu\nu} = pg_{\mu\nu}$,
where the indices $\mu, \nu,$ run over spacelike values {1,2,3}. 

\begin{equation}
T^{\mu}_{\nu} = \left( \begin{array}{ccccc}   
\rho &0   &0  &0 \\
0    &-p  &0  &0 \\
0    &0   &-p &0  \\
0    &0   &0  &-p
\end{array} \right)
\label{sten}
\end{equation}

The conservation of energy equation, $\bigtriangledown_{\mu}T^{\mu\nu}= 0$ then implies $\rho \propto a^{-n}$, with $n = 3(1+\omega)$.

Some examples of equations of state are:

\begin{equation} 
\begin{array}{ccccc}
\rho \propto a^{-3} &\leftrightarrow &p = 0  &\leftrightarrow &matter, \\
\rho \propto a^{-4} &\leftrightarrow  &p = \frac{1}{3}\rho  &\leftrightarrow &radiation, \\
\rho \propto a^{-3} &\leftrightarrow &p = -\rho &\leftrightarrow &vacuum
\end{array}
\label{sten2}
\end{equation}

The vacuum energy density, equivalent to a cosmological constant $\Lambda$ via $\rho_{\Lambda} = \Lambda/8\pi G$, is by definition the energy remaining when all other forms of energy and momentum have been cleared away.  

An expanding and cooling Universe leads to a number of predictions: the formation of nuclei and the resulting primordial abundances of elements, and the later formation of neutral atoms and the consequent presence of a cosmic background of photons, the cosmic microwave background (CMB). 
A clear picture of how the Universe evolved to this present time is given as:

\begin{itemize}
\item T $\thicksim 10^{15}K$, t $\thicksim 10^{-12}$ sec: Primordial soup of fundamental particles
\item T $\thicksim 10^{13}K$, t $\thicksim 10^{-6}$ sec: Protons and neutrons form.
\item T $\thicksim 10^{10}K$, t $\thicksim 3 min$: Nucleosynthesis: nuclei form.
\item T $\thicksim 3000 K$, t $\thicksim$ 300,000 years: Atoms form (emission of CMB).
\item T $\thicksim 10K$, t $\thicksim 10^{9}$ sec: Galaxies form.
\item T $\thicksim 3K$, t $\thicksim 10^{10}$ years: Today.
\end{itemize}

\subsection{Big/Bang Model}  

The Big-Bang Model of the Universe explains that the Universe started as a hot bowl of soup from a singularity with infinite density and temperature and it is expanding and cooling with time. This Model is based on the observation of cosmic microwave background radiation (CMB)$~\cite{cmb}$ and the observed expansion of the Universe by Edwin Hubble. Measurement of CMB also suggest a flat Universe ($\Omega = 1$). 

Observation of the first acoustic peak, first accomplished with precision by the Boomerang$~\cite{boom}$ and MAXIMA$~\cite{max}$ experiments, indicate that the geometry of the Universe is flat, with $\Omega_{total} = 1.02 \pm 0.05$$~\cite{flatuni}$.
 
However this success of the standard Big Bang leaves us with a number of disturbing puzzles. In particular, how did the Universe get so big, so flat, and so uniform?. This is the flatness and horizon problem. These observed characteristics of the Universe are poorly explained by the standard Big Bang model and needed something to make it fit the data: 
This new model is called Inflation, discovered by Alan Guth in 1980$~\cite{guth}$. 

\subsubsection{Inflation}

Within the context of a standard matter or radiation dominated Universe, the flatness and horizon problems have no solutions simply because gravity will curve spacetime causing a decelerating in the expansion and an eventual collapse to a singularity. Though invoking initial conditions that the Universe started out flat, hot and in thermal equilibrium, could resolve the horizon and flatness problem, it is not a satisfactory explanation. There need to be an explanation of why these initial conditions existed. This explanation was found by Alan Guth in his Inflational model. Inflation is the idea that at some very early epoch, the expansion of the Universe was accelerating instead of decelerating. 
It is evident from the Friedman equation below that the condition for acceleration $\ddot{a} > 0$ is that the equation of state be characterized by negative pressure, $1+3\omega < 0$.

\begin{equation}
\frac{\ddot{a}}{a} = -(1+3\omega)\left( \frac{4\pi G}{3}\rho \right). \label{inflaeq1} 
\end{equation}

This means that the Universe evolves toward flatness rather than away.
In accelerated expansion, physical distance d between the two points increases linearly with the scale factor, $d\propto a(t)$.
The horizon size\footnote{The horizon size of the Universe is how far a photon can have traveled since the Big Bang} is proportional to the  inverse of the Hubble parameter $d_{H} \propto H^{-1}$.
This implies that two points that are initially in casual contact $(d < d_{H})$ will expand so rapidly that they will eventually be casually disconnected.
Thus accelerating expansion provides a tangible resolution to the horizon and flatness problems.
The Inflationary theory thus remains up to date the best theory for cosmic evolution.

\section{The Cosmological Constant}

General relativity together with cosmological principle leading to the 
Friedman equations ($~\ref{cosmoeq3}$ $~\ref{cosmoeq4}$), shows that space-time is dynamic and the Universe is not static but either expanding or contracting. 
In order to get a static Universe to match with reality\footnote{It was thought at that time that the Universe was static}, Einstein introduced the cosmological constant to modify his field equation ($~\ref{cosmoeq1a}$). This cosmological constant provides an opposing force to gravity and thus holding the Universe closed and static. He arrived at this by noticing that by adding a constant $\Lambda$ to the stress energy tensor $T_{\mu\nu}$, the conservation equation 

\begin{equation}
D_{\mu}T^{\mu\nu} = 0, \label{cceq1}
\end{equation}

which is an equivalence to charge conservation in electromagnetism, 

\begin{equation}
\partial_{\mu}J^{\mu} = 0, \label{cceq2}
\end{equation}

remains invariant, i.e:

\begin{equation}
D_{\mu}T^{\mu\nu} = D_{\mu}(T^{\mu\nu} + \Lambda g^{\mu\nu}) = 0. \label{cceq3}
\end{equation}

For a homogeneous fluid the stress energy tensor is given by eq.$~\ref{sten}$,
and stress energy conservation takes the form of the continuity:$\footnote{The continuity equation relates the evolution of the energy density to its equation of state $p =\omega \rho$.}$ $~\cite{cosmoinfla}$,

\begin{equation}
\frac{d\rho}{dt} + 3H(\rho + p) = 0. \label{cceq4}  
\end{equation}

Adding a constant term $\Lambda$ to the field equation implies adding a constant energy density to the Universe, which from the continuity equation, eq.$~\ref{cceq4}$, implies negative pressure, $p_{\Lambda} = -\rho_{\Lambda}$.

The Einstein field equation in presence of matter then becomes:

\begin{equation}
R_{\mu\nu} - \frac{1}{2}g_{\mu\nu}R + \Lambda g_{\mu\nu} = 8\pi GT_{\mu\nu}, \label{cceq6}
\end{equation}
 
and the action from which the field equation can be deduced is given by the Einstein-Hilbert action:

\begin{equation}
S = \frac{1}{16\pi G}\int d^{4}x \sqrt{-g}(R - 2\Lambda) + S_{M} \label{cceq7}
\end{equation}

where g is the determinant of the metric tensor $g_{\mu\nu}$, and $S_{M}$ is the action due to matter.

The Friedman equations are then modified to be:

\begin{equation}
H^{2} = \frac{8\pi G}{3}\rho + \frac{\Lambda}{3} - \frac{k}{a^{2}} \label{cceq7a}
\end{equation}

and 
\begin{equation}
\frac{\ddot{a}}{a} = -\frac{4\pi G}{3}(\rho + 3p) + \frac{\Lambda}{3}  \label{cceq7b}
\end{equation}

These equations admit a static solution as Einstein wanted, called ``Einstein static Universe'' with positive spatial curvature k =+1, and all parameters $\rho$, p, and $\Lambda$ nonnegative$~\cite{caroll}$.

Discovery of an expansion of the Universe by astronomer Edwin Hubble in 1920 showed that the Universe is not static, implying no need for a cosmological constant. This is what Einstein famously called his ``greatest blunder''.  
As Einstein also famously quoted ``If there is no quasi-static world, then away with the cosmological term''.
Classical it is okay to remove the cosmological term but quantum mechanically it is difficult due to quantum corrections. Anything that contributes to the energy density of the vacuum acts just like a cosmological constant.

Thus the cosmological constant turns out to be a measure of the energy density of the vacuum (the state with lowest energy). 
The energy-momentum tensor of the vacuum is given by:

\begin{equation}
T_{\mu\nu}^{vac} = -\rho_{vac}g_{\mu\nu}, \label{cceq8}
\end{equation}

where its energy density $\rho_{vac}$, is related to the cosmological constant by:

\begin{equation}
\rho_{vac} = \rho_{\Lambda_{vac}} \equiv \frac{\Lambda_{vac}}{8\pi G}. \label{cceq9}
\end{equation}

Thus the effective cosmological constant $\Lambda_{eff}$ given by:

\begin{equation}
\Lambda_{eff} = \Lambda_{b} + \Lambda_{vac} \label{cceq10}
\end{equation}

 where $\Lambda_{b}$ is the bare cosmological constant and  $\Lambda_{vac} = \rho_{vac}8\pi G$ is the cosmological constant due to the vacuum.

The question now is; where does this vacuum energy or zero point energy come from?.
Quantum mechanics predicts the existence of what are usually called ''zero-point'' energies for the strong, the weak and the electromagnetic interactions, where ''zero-point'' refers to the energy of the system at temperature T=0, or the lowest quantized energy level of a quantum mechanical system. 
In conventional quantum physics, the origin of zero-point energy is the Heisenberg uncertainty principle, which states that, for a moving particle such as an electron, the more precisely one measures the position, the less exact the best possible measurement of its momentum, and vice versa. This minimum uncertainty is not due to any correctable flaws in measurement, but rather reflects an intrinsic quantum fuzziness in the very nature of energy and matter springing from the wave nature of the various quantum fields. This leads to the concept of zero-point energy.
Zero-point energy is the energy that remains when all other energy is removed from a system. These vacuum fluctuations are real and demonstrated by the Casimir effect$~\cite{casimir}$.

How do we then calculate this energy?: A free quantum field can be thought of as a collection of an infinite number of harmonic oscillators, with Hamiltonian

\begin{equation}
H = \hbar\omega \left( \hat{a}^{\dagger}\hat{a} + \frac{1}{2} \right), \label{hamil}
\end{equation}

where $\hat{a}$ and $\hat{a}^{\dagger}$ are the lowering and raising operators, respectively, with commutation relation $[\hat{a},\hat{a}^{\dagger}]=1.$
The results in a stack of energy eigenstates $|n>$:

\begin{equation}
H|n> = \hbar\omega\left( n + \frac{1}{2}\right)|n> = E_{n}|n>.\label{hami2} 
\end{equation} 

The ground state $|0>$ is called the vacuum or zero-particle state.
The ground state energy of a harmonic oscillator is $E_{0}=(1/2)\hbar\omega$
but the ground state energy of the quantum field is a collection of ground state energies of harmonic oscillators as stated above and therefore given by:

\begin{eqnarray}
H|0> & = & \int^{\infty}_{-\infty}d^{3}k\left[ \hbar\omega_{k}\left(\hat{a}^{\dagger}\hat{a} + \frac{1}{2}\right) \right]|0> \\
       &= & \left[ \int^{\infty}{-\infty}d^{3}k(\hbar\omega_{k}/2) \right]|0> \\
       &= \infty   \label{hami3}  
\end{eqnarray}

where the index k is the momentum, and $\omega_{k}$ is the angular frequency at that momentum.
Thus we see that the ground state energy diverges. However, if one should assume that general relativity holds true up to the Planck scale, $m_{PI}\thicksim 10^{19} GeV$, one may introduce a momentum cutoff at this scale. The ground state energy then becomes finite. The cosmological constant $\Lambda_{vac}$ for the vacuum can then can then be taken to be $\Lambda\approx (8\pi G)^{-1/2}$, G is Newton's gravitational constant. This gives the energy density of the vacuum to be:

\begin{equation}
\rho_{vac} \thickapprox 2^{-10}\pi^{-4}G^{-2} = 2 \times 10^{71} GeV^{4}
\label{hami4} 
\end{equation}

Recent astronomical evidence starting in 1997 Type I Supernovae$~\cite{sup}$,
 WMAP$~\cite{wmap}$, Boomerang$~\cite{boom}$, and SdSS$~\cite{sdss}$  shows a flat Universe with a positive but very small cosmological constant. The evidence shows that the Universe is evolving towards a pure de Sitter space-time with matter energy being diluted with time and the energy density being dominated by the vacuum, $\Omega_{vac} \approx 0.7\Omega$, and $\Omega_{matter} \approx 0.3\Omega$.
 The implication is that of an accelerating expansion. The measured value of this effective vacuum energy density is: 

\begin{equation}
\rho_{eff} = \rho_{vac} + \rho_{b} = 10^{-47}GeV^{4},  \label{hami5} 
\end{equation}

where $\rho_{b} = \Lambda_{0}/8\pi G$, and $\Lambda_{b}$ is the bare cosmological constant. Thus there is a huge discrepancy of about 120 orders of magnitude between the measured energy density and the theoretical prediction.
Note that even if we set the cut off scale at 1 TeV (the supersymmetry breaking scale) the difference is still huge; of 59 orders of magnitude. And even if we only worry about zero-point energies in quantum chromodynamics, we would expect $\rho_{vac}$ to be of order $\Lambda^{4}_{QCD}/16\pi^{2}$, or $10^{-6}GeV^{4}$, requiring $\Lambda_{b}/8\pi G$ to cancel this term to about 41 decimal places$~\cite{weinb}$. An incredible amount of fine-tuning.
 
 This is the ``Cosmological Constant Problem'', currently the most challenging issue in physics yet to be resolved. Not only do we want to understand why the cosmological constant is small but also why it is not exactly equal to zero and why is its energy density today of about the same order of magnitude as the matter energy density. 
Thus what Einstein called his greatest blunder has turned out to be his greatest insight.

A number of attempts have been made at resolving the cosmological constant puzzle. However no clear cut solution has yet been found. 
The approaches to resolving this puzzle can be divided into five main categories$~\cite{ste}$:

\begin{itemize}
\item  Fine-tuning
\item  Symmetry, e.g Supersymmetry
\item  Back-reaction Mechanism e.g., Quintessence scalar field
\item  Violating Equivalence Principle e.g., Non-local Gravity, Massive Gravitons
\item  Statistical Approaches e.g., Anthropic Principle, Quantum cosmology, $\Lambda-N$ correspondence, Wormholes
\end{itemize} 

Here I discuss a few of these attempts:

\subsection{Fine-tuning}

The bare cosmological constant can be adjusted by hand to match the observed data. However this will involve an incredible amount of precision tuning. Assuming we set the energy scale at I TeV, a precision tuning to an order of 59 decimal places is needed. For the Planck scale a tuning of order of 120 decimal places is needed. The smallest deviation from this tuning will affect structure formation in the Universe. This tuning method is thus not well accepted in Physics.

 \subsection{Supersymmetry}

Supersymmetry (SUSY) is a space-time symmetry relating bosons to fermions. 
Supersymmetry is associated with ``supercharges'' $Q_{\alpha}$, where $\alpha$ is a spinor index$~\cite{ssy, ssyp}$. This is analogous to ordinary symmetries which are associated with conserved charges.

I begin by looking at globally supersymmetric theories. 
In SUSY, the Hamiltonian is related to the supercharges by:

\begin{equation}
H = \sum_{\alpha}{Q_{\alpha},Q_{\alpha}^{\dagger}}, \label{susy}
\end{equation}

In a completely supersymmetric state in which $Q_{\alpha}|\psi> = 0$, for all $\alpha$, i.e the vacuum state, the Hamiltonian vanishes, $<\psi|H|\psi> = 0$$~\cite{ssypp}$. 

To calculate the effective energy of the vacuum, we need to sum energy from vacuum fluctuations and that of a scalar potential V\footnote{A scalar field has the same symmetries as that of the vacuum}.

In supersymmetry we expect equal number of bosons and fermions. The quantum corrections to the vacuum coming from bosons are of the same magnitude but opposite sign compared to that of the fermions and the two effects cancel each other.
Thus in supersymmetric theories the energy from vacuum fluctuations is zero.
For the scalar field, the potential is given as a function of the superpotential $W(\phi^{i})$ and is given by:

\begin{equation}
V(\phi^{i},\overline{\phi}^{j}) = \sum_{i}|\partial_{i}W|^{2} \label{ssyeq}
\end{equation}

where $\partial_{i}W = \partial W/\partial \phi^{i}.$ Unbroken SUSY only occurs for values of $\phi^{i}$ such that $\partial_{i}W = 0$, implying $V(\phi^{i},\overline{\phi}^{j}) = 0$. Thus we can deduce that the effective vacuum energy of a supersymmetric state in a globally supersymmetric theory vanishes.

However in supergravity theories the scalar field potential V, does not only depend on the superpotential $(W(\phi^{i})$, but also on a ``Kahler potential'', $K(\phi^{i},\overline{\phi}^{j})$, and the Kahler metric $K_{ij}$  constructed from the Kahler potential by $K_{ij} = \partial^{2}K/\partial\phi^{i}\partial \overline{\phi}^{j}$. The scalar potential is given by$~\cite{caroll}$:

\begin{equation}
V(\phi^{i},\overline{\phi}^{j}) = e^{K/M^{2}_{PI}}\left[K^{i\overline{j}}
(D_{i}W)(D_{\overline{j}}\overline{W}) - 3M_{PI}^{-2}|W|^{2},   \right] \label{ssyeq2}
\end{equation}

where $D_{i}W$ is the Kahler derivative:

\begin{equation}
D_{i}W = \partial_{i}W + M_{PI}^{-2}(\partial_{i}K)W. \label{ssyeq3}
\end{equation}

In an unbroken supersymmetry the Kahler derivative is zero hence the potential is negative, and thus the effective cosmological constant is negative when gravity is added in supersymmetric theories.

We can then conclude that the vacuum state in an unbroken supersymmetric theory has zero or negative energy, $\Lambda \leq 0$.

In summary, if our world is supersymmetric then the vacuum energy will be expected to be zero or negative. 
However no supersymmetric partners of the Standard Model particles has yet been found and hence we expect supersymmetry to be a broken symmetry below 1 TeV, implying a large vacuum energy. Thus supersymmetry does not help us in solving the cosmological constant problem.

There is however a nice suggestion by Witten$~\cite{wit}$ that in a 2+1 space-time dimensions one can have supersymmetry of the vacuum (and thus $\Lambda = 0$) without supersymmetry of the spectrum i.e no Bose-Fermi degeneracy for particles.   If this is true, the implication is that we can find a nonsupersymmetric string vacua with zero cosmological constant. 

 \subsection{Quintessence}
  
The idea of quintessence is that the acceleration of the Universe is driven by a dynamic field, a scalar whose value slowly changes with time. Thus the cosmological constant is small because the Universe is old. This idea of a scalar field driving expansion fits perfectly well with inflation since we expect inflation to stop at some point which will imply a vacuum energy with a constant value over all time will not be ideal in explaining cosmic evolution. The value of this scalar field is expected to be high at the time of inflation and has reduce to a constant small value at this present epoch but still rolling down to its equilibrium vacuum point.

In a homogeneous Universe a scalar field is a function of time only. One can imagine a a uniform scalar field $\phi(t)$ rolling down a potential $V(\phi)$ with
energy density given by:

\begin{equation}
\rho_{\phi} = \frac{1}{2}\dot{\phi}^{2} + V(\phi) \label{ener}
\end{equation} 

and pressure given by:

\begin{equation}
p_{\phi} = \frac{1}{2}\dot{\phi}^{2} - V(\phi). \label{pres}
\end{equation}

The first terms in both equations are the kinetic energy terms and the second term is the potential energy.

There are various choices of the form of this potential corresponding to different models of inflation, some of which are:

\begin{equation}
V(\phi) = \lambda(\phi^{2} - M)^{2} \  Higgs \ potential \label{higss}
\end{equation}

\begin{equation}
V(\phi) = \frac{1}{2}m^{2}\phi^{2} \  Massive \ scalar \ field \label{hs}
\end{equation}

\begin{equation}
V(\phi) = \lambda \phi^{4} \  Self-interacting  \ scalar \ field \label{hs1}
\end{equation}

Substituting eqs.($~\ref{ener}$\&$~\ref{pres}$) into the Friedman and fluid equations eqs.($~\ref{cosmoeq3}$\&$~\ref{cceq4}$) respectively gives an expression for the rate of expansion:

\begin{equation}
H^{2} = \frac{8\pi}{3m^{2}_{PI}}\left[ V(\phi) + \frac{1}{2}\dot{\phi}^{2} - \frac{k}{a^{2}} \right],  \label{quin}
\end{equation}

and equation of motion:

\begin{equation}
\ddot{\phi} + 3H\dot{\phi} + V^{\prime}(\phi) = 0 \label{quin2}
\end{equation}

From the second Friedman equation we can see that:

\begin{equation}
\ddot{a} > 0 \Longleftrightarrow p < -\frac{\rho}{3} \Longleftrightarrow \dot{\phi}^{2} < V(\phi). \label{infl}
\end{equation}

This implies that inflation (accelerated expansion) starts when the potential energy of the scalar field dominates its kinetic energy.  The potential must be chosen such that it is flat enough for the scalar field to roll slowly and have a minimum at which inflation can end. This strategy for choosing the potential to behave this way is called the {\it slow-roll approximation}.
Since the potential dominates, eqns ($~\ref{quin}$) and ($~\ref{quin2}$) can be approximated to:

\begin{equation}
H^{2} \approx \frac{8\pi}{3m^{2}_{PI}}V \label{quin3}
\end{equation}

\begin{equation}
3H\dot{\phi} \approx -V^{\prime} \label{qn}
\end{equation}

And slow-roll parameters are defined as$~\cite{liddle}$:

\begin{equation}
\epsilon(\phi) = \frac{m^{2}_{PI}}{16\pi}\left(\frac{V^{\prime}}{V}, \right)^{2}
\label{slow}
\end{equation}

\begin{equation}
\eta(\phi) = \frac{m^{2}_{PI}}{8\pi}\frac{V\prime\prime}{V},
\label{slow2}
\end{equation}

$\epsilon(\phi)$ measures the slope of the potential and $\eta(\phi)$ measures the curvature. The necessary conditions for slow-roll approximation to hold are:

\begin{equation}
\epsilon(\phi) \ll 1; \ |\eta| \ll 1 \label{slow3}
\end{equation}

The problem, of course is to explain why $V(\phi)$ is small or zero at the value of $\phi$ where $V^{\prime}(\phi) = 0$. Some compelling explanations have been given by 'tracker' methods$~\cite{tracker}$. 

\subsection{Violation of the Equivalence Principle} 

The equivalence principle states that gravity couples to all forms of energy. Since high cosmological constant value is expected than observed in the curvature and acceleration of the Universe, it is speculated that probably vacuum energy in contrary to ordinary matter does not gravitate, a violation of the equivalence principle. This situation would require modification of general relativity. The vacuum energy would then decouple from gravity, and its relevance will be eliminated and there would be no need to worry about the cosmological constant. 
It is also speculated that the large missen vacuum energy could possibly be the cause of curvature of the extra dimensions. In this case there is no violation of the equivalent principle\footnote{No violation of the equivalence principle has yet been found from experiments}. 
An example of this is found in Braneworld models such as the Randall-Sundrum models which is discussed below:

\subsection{ Randall-Sundrum Models, Warped Extra Dimensions}

The general idea is that our world is confined on a hypersurface, a brane, embedded in a higher dimensional space-time. The standard model fields are restricted to live on a 3-brane, while only gravitons can propagate in the full higher dimensional space.

There are two Randall-Sundrum models, RS-I and RS-II.  I will start with the  RS-I model. In this model there are two branes 3-branes located at separate distances in a five dimensional space which is a foliation with four  dimensional Minkowski slices. The fifth dimension is compactified on an orbifold $S^{1}/Z_{2}$, and the 3-branes are located at the orbifold fixed planes $(at \phi = 0 and \phi = \pi)$, where $\phi$ is the extra fifth dimension called the bulk.

One brane called the ``hidden brane'' has positive tension, while the other one, the ``visible brane'', on which we are supposed to live, has negative tension. Both could have gauge theories living on them. All of the standard model fields are localized on the brane, and only gravity can propagate through the entire higher dimensional space.

The action is given by$~\cite{4}$:

\begin{equation}
S = S_{bulk} + S_{vis} + S_{hid} \label{bact} 
\end{equation}

$S_{b1}$ and $S_{b1}$ are the actions on the branes. 
The action in the bulk space is given by:

\begin{equation}
S_{bulk} = \int d^{4}x \int^{\pi}_{-\pi} d\phi\sqrt{-G}(2M^{3}R - \Lambda), \label{bact2} 
\end{equation}

$\Lambda$ is the bulk cosmological constant and $M$, and $G_{MN}$ are the Planck mass and metric in five dimensions respectively.

The induced metrics on the branes are then given by:

\begin{equation}
g_{\mu\nu}^{hid} = G_{\mu\nu |\phi=0} \ ,g_{\mu\nu}^{vis} = G_{\mu\nu |\phi=\pi}\label{bact3}. 
\end{equation}

where $\mu,\nu$ = 0,...3.

We assume that the fields being localized on the brane are in the trivial vacuum and take into account only nonzero vacuum energies on the branes.

Identifying nonzero vacuum energies on the brane as tensions on the brane, $T_{hid}$ and $T_{vis}$, the brane actions read

\begin{equation}
S_{hid} + S_{vis} = -\int d^{4}x (T_{hid}\sqrt{-g^{hid}} + T_{vis}\sqrt{-g^{vis}},\label{bact4}  
\end{equation}

From the above action the equation of motion becomes:

\begin{equation}
M_{p}\sqrt{G}\left( R_{MN} - \frac{1}{2}G_{MN}R\right) = M_{p}\Lambda\sqrt{G}G_{MN} + 
T_{hid}\sqrt{g_{hid}}g_{\mu\nu}^{hid}\delta_{M}^{\mu}\delta_{N}^{\nu}\delta(\phi) + T_{vis}\sqrt{g_{vis}}g_{\mu\nu}^{vis}\delta_{M}^{\mu}\delta_{N}^{\nu}\delta(\phi - \pi) \label{bact5}
\end{equation}

where the indices M,N = 0,..4, and $M_{p}$ is the five dimensional Planck mass, which has to satisfy $M \gtrsim 10^{8}GeV$, in order not to spoil Newtonian gravity at distances $l \lesssim 0.1 mm$.
With the above assumptions for the brane-tensions and bulk CC, it can be shown that there exists the following static solution, with a flat 4D-metric:

\begin{equation}
ds^{2} = \exp^{-|\phi|/L}\eta_{\mu\nu}dx^{\mu}dx^{\nu} + d\phi^{2} \label{bact6}
\end{equation}

The warp factor in the above equation leads to suppression of all masses on the visible brane in comparison to their natural value. This is used as an explanation for the hierarchy problem. An example is given by the higgs mass:

\begin{equation}
m^{2} = \exp^{-\phi_{0}/L}m_{0}^{2} \label{higeq1} 
\end{equation}

a small hierarchy in $\phi_{0}/L$ results in a large hierarchy between m and $m_{0}$, resolving the hierarchy problem.

The second model is RS-II; here the extra dimensions can be kept large, uncompactified, but warped and there is only one brane. In this scenario the size of the extra dimensions can be infinite, but their volume $\int d\phi\sqrt{G}$, is  still finite. The warp-factor causes the graviton wavefunction to be peaked near the brane, or, in other words, gravity is localized, such that at large 4D-distances ordinary general relativity is recovered. 

The action is given by:

\begin{equation}
S = \frac{1}{2}M_{p}^{3}\int d^{4}x\int^{+\inf}_{-\inf}d\phi\sqrt{G}(R_{5} - 2\Lambda_{5}) + \int d^{4}x\sqrt{g}(\Lambda_{4} + {\it L}_{SM})\label{acct}
\end{equation}

Ignoring ${\it L}$, the equation of motion from extremizing the action with the brane located at $\phi = 0$ is given by:

\begin{equation}
M_{p}\sqrt{G}\left( R_{MN} -\frac{1}{2}G_{MN}R \right) = \\
-M_{p}^{3}\Lambda_{5}\sqrt{G}G_{MN} + \Lambda_{4}\sqrt{g}g_{\mu\nu}\delta_{A}^{\mu}\delta_{B}^{\nu}\delta(\phi), \label{act}
\end{equation}

This equation gives a flat space solution at the expense of fine-tuning $\Lambda_{5}$ and $\Lambda_{4}$.

The problem with RS models is that fine-tuning of the parameters i.e the tensions on the brane is necessary in order to get a flat space solution and an effective vanishing (almost zero) CC in four dimensions. The fine-tuning constraints are given by:

\begin{equation}
T_{hid} = -T_{vis} = 24M_{p}^{3}k, \ k^{2} = -\frac{\Lambda}{24M^{3}_{p}} 
\end{equation}

\subsection{Statistical Approach} 

\subsubsection{Anthropic Principle} 

The idea behind the anthropic principle is that the parameters we call constants of nature may in fact be stochastic variables taking different values in different parts of the Universe, and the values of the parameters we observe are determined by chance and by anthropic selection. Taking the cosmological constant for an example; the value observed by any species of astronomers will be conditioned by the necessity that this value of $\rho_{V}$ should be suitable for the evolution of intelligent life.  There is a range for which some of these constants must lie to support life. This range is called the anthropic bound. Given a parameter X, its range to support life may be given as:

\begin{equation}
X_{min} < X < X_{max}  \label{ant}
\end{equation}

Values of X outside the interval are not going to be observed, because such values are inconsistent with the existence of observers. This is the ``anthropic principle'' $~\cite{weinb},~\cite{vilen}$.

Thus we measure a small value of the cosmological constant because it is the ideal value to support life. Any value higher will not lead to structure formation in our Universe and hence life and we will not be present to observe it.
From string theory point of view anthropic principle will imply that there exist a moduli space of supersymmetric vacua called supermoduli space$~\cite{skind}$. Moving around on this moduli space is accomplished by varying certain dynamic moduli. These moduli are scalar fields which determine the size and shape of the compact internal space, and each have their own equations of motion. Thus there is only one theory but many solutions characterized by the values of the scalar field moduli. The value of the potential energy of the scalar field at the minimum is the cosmological constant for that vacuum. In unbroken supersymmetry, the moduli fields have zero potential at their minima but once supersymmetry is broken there can be several minima. Thus there is a landscape of string vacua with different cosmological constants one of which is our own.

\subsubsection{Quantum cosmology}
Hawking suggested that state vector of the Universe could be taken as a superposition of states with different values $\Lambda_{eff}$ with a huge peak at zero$~\cite{haw}$. 

 The wave-function of the Universe $\Psi$, is obtained by a Euclidean path-integral over all metrics $g_{\mu\nu}$ and matter fields $\phi$, defined on a 4-manifold $M_{4}$$~\cite{weinb}$.

 \begin{equation}
\Psi \propto \int [dg][d\phi]\exp^{(-S[g,\phi])}, \label{euc}
\end{equation}

where S is the Euclidean action given by 

\begin{equation}
S = \frac{1}{16\pi G}\int_{M_{4}}\sqrt{g}(R + 2\Lambda_{eff}) + matter \ terms + surface \ terms
\end{equation}

Different Universes with different values of $\Lambda_{eff}$ contribute to this path integral. The probability P, of observing a given field configuration will be proportional to$~\cite{stefan}$: 

\begin{equation}
  P \propto \exp^{-S(\Lambda_{eff})} \propto \exp^{(3\pi\frac{M_{P}^{2}}{\Lambda_{eff}})} \label{pb2}
\end{equation}

where S is the action taken at its stationary point to evaluate the path integral and $M_{P}$ is the Planck's mass. This probability peaks at $\Lambda_{eff} = 0$.
The issue of whether this really solves the cosmological constant problem has been raised by Weinberg$~\cite{weinb}$.

This Euclidean path integral approach may also help to resolve the issue of not being able to define an S-Matrix in a de Sitter space-time as we will see in the later sections. Hawking's argument was that any consistent theory of gravity should involve an appropriate integral over all topologies(trivial and non-trivial)The trivial one being Minkowski space-time. Since Euclidean path integral over
the non-trivial topologies (eg de Sitter space-time) gives non-unitary contributions and hence information loss, it results to the scattering amplitude decaying exponentially with time. This leads to unitary contributions from only the trivial topologies$~\cite{mav}$. 
   
\subsubsection{$\Lambda$-N correspondence}

A proposal towards the solution of the cosmological constant problem was made by Banks$~\cite{banks}$ who conjectured that the cosmological constant should not be viewed as an effective parameter to be derived in a theoretical framework like quantum field theory or string theory, but instead determined as the inverse of the number of degrees of freedom, N, in the fundamental theory. 

The argument towards this approach was laid as follows-:

A black hole has an event horizon and its entropy is bounded by the area of the event horizon given by Bernstein-Hawking's formula:

\begin{equation}
S= \frac{1}{4G_{N}}A \label{ent}
\end{equation}

where A is the surface area, $G_{N}$ is Newton's gravitational constant    

and a temperature given by:
\begin{equation}
T_{H} = \frac{\hbar}{8\pi M}
\end{equation}

where M is the mass of the black hole. Any entropy calculation inside the black hole should be bounded above by the area of the surface. 
Information on the surface of the black hole is a projection of what is going on inside the black hole, a form of holography or what is called UV-correspondence. 
Quantum mechanical calculation of the entropy given by:

\begin{equation}
S \sim \ln(N)  \label{micro}
\end{equation}

where N is the number of microstates or degrees of freedom,
gives a good agreement with the classical value in $~\ref{ent}$.
This implies a bound on the number of degrees of freedom.
 
A dS space-time has a cosmological event horizon. Any
observer in an asymptotically dS space (AsdS) only sees a finite portion of the Universe bounded by a cosmological event horizon. Hence making an analogy between black hole and a dS space-time we may deduce that a dS has a finite entropy given by the above relationship. This implies entropy or number of degrees of freedom in a dS space-time is bounded by the area of the horizon. Thus an asymptotically de Sitter Universe can be described by a finite number of states, given by the Bernstein-Hawking formula stated above.

Now relating the cosmological constant to N, implies a bound on the cosmological constant. This is the central idea of Banks proposal which states that:

``The Bernstein-Hawking Entropy of Asymptotically de Sitter (AsDS) spaces represents the logarithm of the total number of quantum states necessary to describe such a Universe. This implies that the cosmological constant is an input to the theory rather than a quantity to be calculated '' $~\cite{banks}$.    

We know that the structure of AsdS Universe automatically breaks SUSY. The expected SUSY breaking scale is given by:

\begin{equation}
M_{SUSY} \thicksim M_{P}(\Lambda/M_{P}^{4})^{\alpha} \label{ssyr}
\end{equation} 

where $\Lambda$ is the effective cosmological constant and $M_{P}$ is Planck's mass. There exist a good agreement between the observed SUSY breaking scale from experiment and the value obtained from $eq.~\ref{susy}$ provided the value of the exponent $\alpha$ is 1/8 rather than its classical value 1/4. This leads to the question of why isn't the scale of SUSY breaking related to the cosmological constant by the standard classical SUGRA formula, which (without fine tuning) predicts $M_{SUSY} \thicksim \Lambda^{1/4}$.  A suggestion by Banks is that this is attributed to the effect of large virtual black holes via the UV/IR correspondence in M-Theory$~\cite{banks}$.

A summary of this conjecture stated by Banks as follows:

``The structure of an AsdS
Universe automatically breaks supersymmetry (SUSY). From this point of view, the ``cosmological constant problem'' is the problem of explaining why the SUSY breaking is so much larger than that associated in classical supergravity (SUGRA) with the observed value of (bound on ?) the cosmological constant.''    

Summarizing in other words: In the presence of a cosmological constant $\Lambda$, the Universe tends to evolve towards a pure de Sitter space\footnote{ Note that matter and radiation density $\rho_{m} \propto a^{-3}$, $\rho_{m} \propto a^{-4}$ respectively and thus the matter energy density gets diluted and tending towards zero as the Universe expands whilst the vacuum energy density stay constant. Thus we see the Universe becoming asymptotically de Sitter with time.} 
with finite entropy $S = N = 3\pi/\Lambda$, given by the area of the cosmic horizon.
This $\Lambda-N$ correspondence asserts that a Universe with positive $\Lambda$ has degrees of freedom $N = 3\pi/\Lambda$. But to make this theory testable a bound is placed on the number of degrees of freedom.   Thus the entropy an AsdS Universe cannot exceed $N = 3\pi/\Lambda$, called the N-bound.

$\Lambda-N$ correspondence does not solve the cosmological constant problem but rather gives a new insight in physics given as the N-bound on the entropy of a positive $\Lambda$ Universe and its inverse relation to $\Lambda$ $~\cite{bousso}$.

\section{String Theory \& Cosmology}

An accelerating Universe implies a de Sitter space-time. A de Sitter space-time will have serious implications for string theory. The reason being that the only well defined observable in string theory is the S-Matrix which is defined from negative infinity to positive infinity in flat space-time.
However a de Sitter space-time has a cosmic horizon and the killing vectors are not globally defined over all space-time. They are only defined within the horizon. This implies a violation of unitarity and S-Matrix cannot be defined. Let me digress a little on it or look at at in another perspective:
For a black hole, an asymptotic observer lying far outside the horizon will see thermal radiation coming from the horizon. Information coming out as thermal radiation from the black hole is mangled leading to quantum decoherence and thus we have Hawking's ``Information lost paradox''.  In this scenario, an initially pure quantum state will be observed as a mixed state by the asymptotic observer. Unitariy is violated and S-Matrix is not well defined. This situation is similar to the dual dS case except the observer is inside the cosmic horizon and is immersed in a thermal bath of quantum radiation from the cosmic horizon. This is the situation we face developing a quantum gravity theory in a de Sitter space-time
\footnote{There is non-unitarity evolution of quantum states of matter in Black hole and dS space-time}

\begin{center}
\includegraphics[width=6.5in]{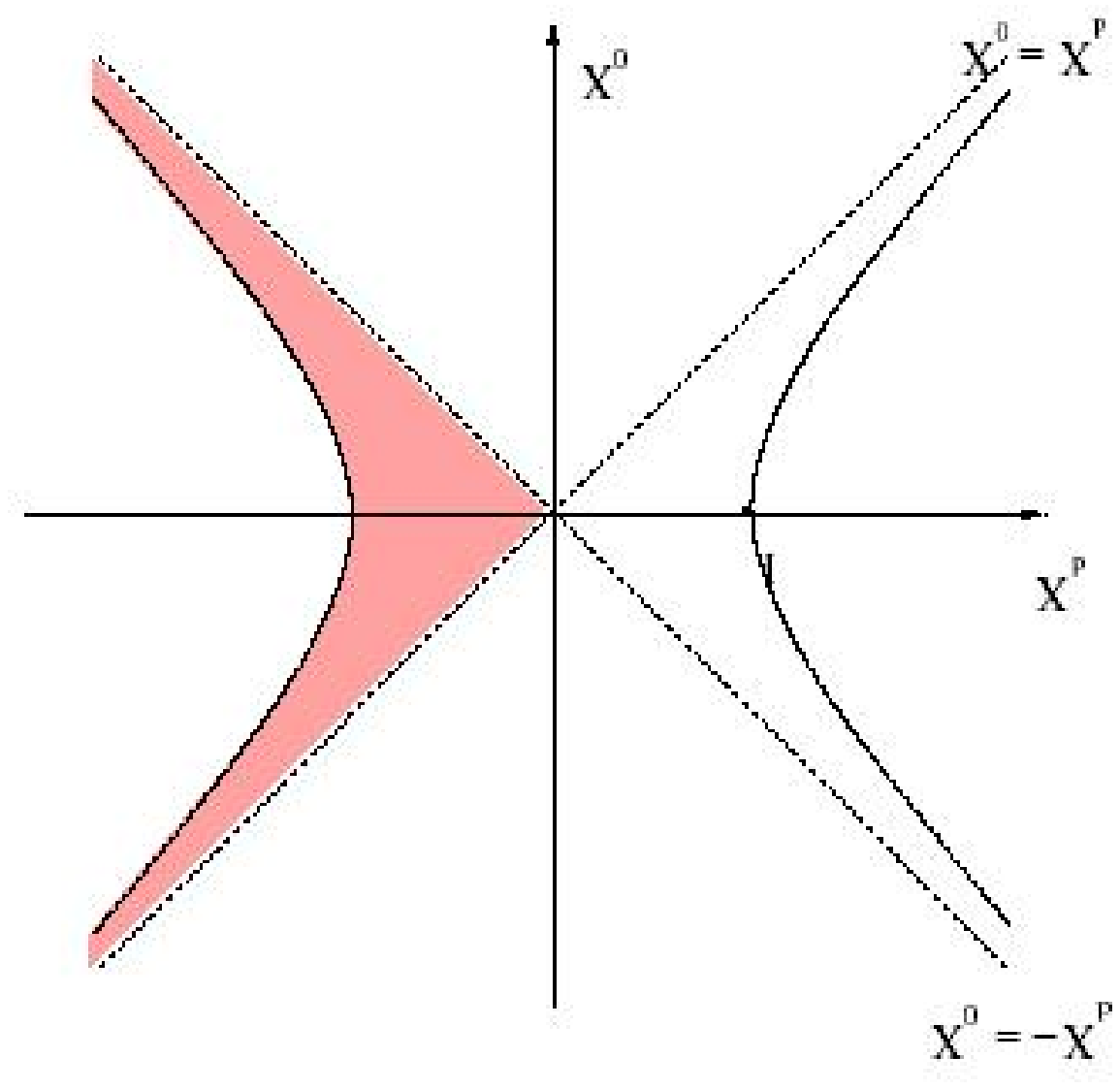} 
{\it  Figure   The region bounded by the de Sitter horizon $(r\leq l)$ in static coordinates is shown by shaded region}
\label{12}
\end{center} 

Three suggested ways out of this problem are:

\begin{itemize}

\item The Quintessence scenario discussed above: 

The implication is that the rolling scalar field has not yet reached its equilibrium energy which is expected to be zero. Implying a vanishing of the cosmic horizon at the vacuum point and S-Matrix will then be defined. 

\item dS/CFT Correspondence:

An S-Matrix is defined in AdS space-time. The AdS/CFT correspondence, a holographic principle conjecture by Maldecena, relates the CFT theory on the boundary of AdS space-time to the gravity theory. Boundary correlators in AdS space-time have been found to relate with the correlators of local gauge invariant conformal field theory (CFT) fields. But our Universe is not AdS (negative cosmological constant) but dS, based on current evidence. A dS/CFT correspondence by $~\cite{dscft}$ has shown relation of correlation functions of string theory in dS space-time and CFT. This holographic principle, averts the ``Information loss paradox'' leading in a big step to a formulation of a quantum theory of gravity. But it may not be valid for the non-conformal and non-supersymmetric case, which is the realistic one we want for our Universe $~\cite{mav}$.

Then then leads us to finding a framework for defining strings in non-conformal backgrounds, such as our dS Universe. The theory behind this is called non-critical or Liouville string.
 
\item Noncritical (Liouville) string Framework

The main idea behind noncritical (Liouville) string framework is identifying the Liouville mode with target time. By doing so, non-conformal backgrounds and positive $\Lambda$ can be accommodated in string theory. I will not discuss this framework here. See $~\cite{ellis}$ for more details.   

\end{itemize}

Conventional big bang cosmology has not yet produced a satisfactory explanation of the small value of the cosmological constant. An attempt by String/M Theory in this direction is given by the Cyclic model. I give a brief discussion of this model below.

\section{Ekpyrotic/Cyclic Model} 

In this model the Universe undergoes a periodic sequence of expansion and contraction. Each cycle begins with the Universe expanding from a ``big bang'', a phase which involves a period of  radiation, matter and dark energy(quintessence) domination, followed by an extended period of cosmic acceleration at low energies which is the current epoch we are observing before finally contracting to a big crunch. The cycle then begins again after the big crunch. 
The cosmic acceleration phase which follows the radiation and matter dominated phase is a crucial epoch in which the Universe approaches nearly vacuous state and restores itself to the initial vacuum state conditions before each big crunch by removing the entropy, black holes, and other debris produced in the preceding cycle. This allows the cycle to repeat making it an attractor solution.

 This cyclic model is based on ideas drawn from the ekpyrotic model$~\cite{ekp}$ and M-theory. The ekpyrotic model explains the Universe as beginning from collision of branes in a 5D bulk space approaching each other along the fifth dimension.
 According to the ekpyrotic model we live at one of the two heavy 4D branes called boundary branes in a 5D Universe described by the Horava-Witten (HW) theory$~\cite{hw}$, where our brane is called the visible brane and the second brane is called the hidden. This is similar to the prescription in Randal-Sundrum model discussed previously, but in addition there is also a 'light' bulk brane at a distance Y from the visible brane in the 5th direction.   
The bulk brane moves toward our brane and collides with it. The residual kinetic energy carried by the bulk brane before the collision then transforms into radiation which is deposited in the three dimensional space of the visible brane. The visible brane, now filled with hot radiation begins to expand as a flat FRW Universe. In this model the issues of homogeneity, isotropy, flatness, and horizon of the Universe do not appear since the three branes are assumed to be initially in a nearly stable BPS state which is homogeneous.

The cyclic model is an improved version of the ekpyrotic model. The inter-brane distance is parameterized by a four-dimensional scalar field, $\phi$\footnote{The brane separation goes to zero as $\phi$ goes to $-\infty$, and the maximum brane separation is attained at some finite value$\phi_{max}$}.
In this scenario the source of the dark energy or the cosmological constant which causes the cosmic acceleration is the inter-brane potential energy from the scalar field. For more detailed review of the cyclic model see$~\cite{cyc}$.

\section{Conclusions}

Though String/M Theory has not fulfilled all the requirements stated earlier for a full quantum theory of gravity it has made some interesting discoveries that may be useful in the final quantum theory. The AdS/CFT correspondence has shown that it is possible to relate a gauge theory to a gravitational theory, and counting of black hole microstates seems to agree with the classical calculations. 
Increasing evidence has shown that our world is asymptotically de Sitter, leading to problems in defining an S-Matrix.  Attempts have been made to find a dS/CFT correspondence but there is doubt if it is applicable in a realistic case of a non-conformal Universe like our own. However there is a suggested solution by using non-critical (Louisville) strings. The critical issue in this model is the identification of the Liouville mode with target time.

The $\Lambda$-N correspondence by Banks does not solve the cosmological constant problem but offers a new perspective, i.e a bound on the entropy of a dS space-time. If it is true, the implication is that M-theory only arises in the limit where the cosmological constant vanishes and N is infinite.  
   
Cosmology provides a unique opportunity for string/M Theory to justify its claim as the best quantum theory. Attempts to explain the small value of the cosmological constant within conventional big bang cosmology has proven to be very difficult and up to date no clear cut explanation without fine-tuning exist. The Cyclic model from a string theory point of view appears to be very promising despite some criticism by some physicists that it is not an alternate inflation theory but just another inflationary theory.

Not having a natural explanation for the vanishing or extreme smallness of the cosmological constant will remain a key obstacle for any further progress in particle physics. 

Suggestion by Witten that in a 2+1 space-time dimensions one can have supersymmetry of the vacuum (and thus $\Lambda = 0$) without supersymmetry of the spectrum, is a very interesting area to look further into. If it is possible to extend this to four space-time dimensions the cosmological constant puzzle may be resolved. 
Another very interesting approach is the Euclidean path integral suggested by Hawking. If it is true, it will solve both the cosmological constant problem as well as the non-unitarity of S-Matrix problem in de Sitter space-time as ours. Since space-time is dynamic it may be possible that our Universe has evolved or on its way to evolving through all topologies and thus we living and observing\footnote{This certainly runs into the anthropic principle} its asymptotically de Sitter evolution stage. Thus the wavefunction should really be an integral over all topologies. But a rigorous mathematical formulation is needed to support this, and certainly an area worth focusing on.
 
The dualities relations between different string theories on different string vacus (space-time backgrounds) is a strong indication that a background independent theory might exist. Efforts are needed to be made towards this goal, though we may be far from understanding the foundations of this background independent theory\footnote{A background independent theory is an immense task and we may be far from it. However loop quantum gravity is background independent but does not unify the forces. The question is whether unification is really needed, in which case loop quantum gravity should do the trick, or if we should extract some information from loop quantum gravity into string theory, or whether we need to merge string theory with it to get the final theory. I have no clue to this.  Only time and some serious efforts will tell.}. 

These efforts towards unification though haven't yet proved successful has generated a lot of insights in physics such as, the holographic principle, the N-bound on entropy, etc. These may be useful and form part of the final unified theory when discovered.

\section{Appendix A}

By world-volume and space-time re-parametrization invariance of the theory, we may choose the ``static gauge'' in which the world-volume is aligned with the first p+1 space-time coordinates, leaving 9-p transverse coordinates. This amounts to calling the p+1 brane coordinates $\xi^{a}=x^{a}, a=0,1,...p$. By this choice the full dynamics of the D-brane can be given by the Dirac-Born-field action:

\begin{equation}
S_{DBI}=-\frac{T_{p}}{g_{s}}\int d^{p+1}\xi \sqrt{-det_{0\leqq a,b \leqq p}(
\eta_{ab}+ \delta_{a}x^{m}\delta_{b}x^{m} + 2\pi\alpha 'F_{ab})}  \label{appAeqn1}
\end{equation}

It describes a model of nonlinear electrodynamics on a fluctuating p-brane.

$T_{p}$ is the tension of the p-brane given by:

\begin{equation}
T_{p}= \frac{1}{\sqrt{\alpha^{\prime}}}\frac{1}{(2\pi\sqrt{\alpha^{\prime }})^{p}}  \label{appAeqn2}
\end{equation}

In the case where there are no gauge field on the Dp-brane, so that $F_{ab}\equiv 0$. Then the Dirac-Born-Infeld action reduces to:

\begin{equation}
S_{DBI}(F=0)= -\frac{T_{p}}{g_{s}}\int d^{p+1}\xi \sqrt{-det_{a,b}(-\eta_{\mu\nu}\delta_{a}x^{\mu}\delta_{b}x^{\mu})}   \label{appAeqn3}
\end{equation}

In the slowly varying field approximation, 
the effective action of a Dp-brane for a trivial metric $G_{\mu\nu}$, and antisymmetric tensor background $B_{\mu\nu}$, as well as a constant dilaton field $\phi$, is given by the Dirac-Born-Infeld action$~\cite{dbi}$ 

\begin{equation}
S_{DBI}=-T_{p}\int d^{p+1}\xi e^{-\phi}\sqrt{-det(g_{ab} + B_{ab} + 2\pi\alpha^{,}F_{ab}) } \label{appAeqn4}
\end{equation}

where $g_{ab} = (\delta X^{\mu}/\delta\xi^{a})(\delta X^{\nu}/\delta\xi^{b})G_{\mu\nu}$, etc are the pull-backs of spacetime supergravity fields $G_{\mu\nu}, B_{\mu\nu}$ to the brane, $F_{ab}={\it d_{a}A_{b}-d_{b}A_{a}}$ is the field strength of the gauge fields living on the Dp-brane worldvolume.

Expanding eq.$~\ref{appAeqn1}$ in flat space ($G_{\mu\nu}=\eta_{\mu\nu}, B_{\mu\nu}=0$)  for slowly varying fields to order $F^{4}$, $(\delta x)^{4}$ gives$~\cite{zabo}$:

\begin{equation}
 S_{DBI} = -\frac{T_{p}(2\pi\alpha\prime)^{2}}{4g_{s}}\int d^{p+1}\xi (F_{ab}F^{ab} + \frac{2}{(2\pi\alpha\prime)^{2}}\delta_{a}x^{m}\delta^{a}x_{m}) - \frac{T_{P}}{g_{s}}V_{p+1} + O(F^{4}), \label{appeqn4}
\end{equation}

where $V_{p+1}$ is the (regulated) p-brane world-volume. This is the action for a U(1) gauge theory in (p+1) dimensions with 9-p real scalar fields $x^{m}$.

 Eq.$~\ref{appeqn4}$ can also be obtained by dimensional reduction of U(1) Yang-Mills theory in ten space-time dimensions, which is defined by the action:

\begin{equation}
S_{YM}=-\frac{1}{4g^{2}_{YM}}\int d^{10}x F^{\mu\nu}F_{\mu\nu}  \label{appAeqn5}
\end{equation}

To show that the ten-dimensional gauge theory action eq.$~\ref{appAeqn5}$ reduces to the expansion eq.$~\ref{appeqn4}$ of the Dp-brane world-volume action (up to an irrelevant constant)
we take the fields $A_{a} (a=0,1....p)$\footnote{These are actually the gauge fields that live on the brane} and $A_{m}=\frac{1}{2\pi\alpha^{,}}x^{m}$ (m=p+1,......10.)\footnote{These are the scalars, and describe the transverse fluctuations of the branes i.e the positions of the branes}, to depend only on the p+1 brane coordinates $xi^{a}$, and be independent of the transverse coordinates $x^{p+1}, ....x^{9}$. This requires the identification of the Yang-Mills coupling constant (electric charge) $g_{ym}$ as:

\begin{equation}
g^{2}_{YM}=g_{s}T^{-1}_{p}(2\pi\alpha\prime)^{-2}=\frac{g_{s}}{\sqrt{\alpha\prime}}(2\pi\sqrt{\alpha\prime})^{p-2} \label{appeqn5}.
\end{equation}

Note that above equation are for a single brane giving a U(1) abelian theory.
For multiple D-branes, one can derive a non-abelian U(N) extension 
of the Dirac-Born-Infeld action and incorporating supersymmetry.

The general theory is stated as:

`` The low-energy dynamics of N parallel, coincident Dp-branes in flat space is described in static gauge by the dimensional reduction to (p+1)-dimensions of {\it N=1} supersymmetric Yang-Mills theory with gauge group U(N) in ten space-time dimensions.''

The ten dimensional action is given by\footnote{Note that this is supersymmetric and non-abelian compared to the abelian case in $~\ref{appAeqn5}$}:

\begin{equation}
S_{YM}=\frac{1}{4g_{YM}^{2}}\int d^{10}x[T(F_{\mu\nu}F^{\mu\nu}) + 2iTr(\overline{\psi}\Gamma^{\mu}D_{\mu}\psi)], \label{appAeqn6}
\end{equation}

where 
\begin{equation}
F_{\mu\nu}=\delta_{\mu}A_{\nu}-\partial_{\nu}A_{\mu} - i[A_{\mu},A_{\nu}]  \label{appAeqn7}
\end{equation}

is the non-abelian field strength of the U(N) gauge field $A_{\mu}$, and the action of the gauge-covariant derivative ${\it D_{\mu}}$ is defined by

\begin{equation}
{\it D_{\mu}=\partial_{\mu}\psi-i[A_{\mu},A_{\nu}]} \label{appAeqn8}
\end{equation}

where $g_{YM}$ is the Yang-Mills coupling constant, $\Gamma^{\mu}$ are $16 \times 16$ Dirac matrices, and the $N \times N$ Hermitian fermion field $\psi$ is a 16-component Majorana-Weyl spinor of the Lorentz group SO(1,9) which transforms under the adjoint representation of the U(N) gauge group. 
After imposition of the Dirac equation (${\it D\psi=\Gamma_{\mu}D_{\mu}\psi}$), 
the field theory eq.$~\ref{appAeqn6}$ possesses eight on-shell bosonic, gauge field degrees of freedom, and eight fermionic degrees of freedom$~\cite{zabo}$.

Let us follow the previous example in the U(1) case to see how to construct a supersymmetric Yang-Mills gauge theory in p+1 dimensions by dimensional reduction from eq.$~\ref{appAeqn6}$. By the same approach, we take all fields to be independent of the coordinates $x^{p+1},....x^{9}$. Then the ten-dimensional gauge field $A_{\mu}$ splits into a (p+1)-dimensional U(N) gauge field $A_{a}$ plus 9-p Hermitian scalar fields $Phi^{m}=\frac{1}{2\pi\alpha '}x^{m}$ in the adjoint representation of U(N). The Dp-brane action is thereby obtained from the dimensionally reduced field theory as:

\begin{equation}
S_{D_{p}} = \frac{T_{p}g_{s}(2\pi\alpha ')^{2}}{4} \int d^{p+1}\xi Tr(F_{ab}F^{ab} + 2{\it D_{a}\Phi^{m}D^{a}\Phi_{m}} + \sum_{m\neq n}[\Phi^{m},\Phi^{n}]^{2} + fermions) \label{appAeqn9}
\end{equation}

where a, b = 0,1......,p, m, n = p+1,....,9 
with no explicit display of the fermionic contributions. 
Thus the brane dynamics is described by a supersymmetric Yang-Mills theory on the Dp-brane world-volume, coupled dynamically to the transverse, adjoint scalar fields $\Phi^{m}$ with $\Phi^{m}$ representing the fluctuations of the branes transverse to their world-volume.

The Yang-Mills potential in eq.$~\ref{appAeqn9}$, which is given by:

\begin{equation}
V(\Phi)=\sum_{m\neq n}Tr[\Phi^{m},\Phi^{n}]^{2} \label{appAeqn10}
\end{equation}

A vacuum solution of the equations of motion requires minimization of the potential. In an unbroken supersymmetric case this potential should be zero. 
Vanishing of the potential in eq.$~\ref{appAeqn10}$, requires the condition: 

\begin{equation}
[\Phi^{m},\Phi^{n}]=0 \label{appAeqn12}
\end{equation}

for all m,n and at each point in the (p+1)-dimensional world-volume of the branes.

This implies that the $N \times N$ Hermitian matrix fields $\Phi^{m}$ are commutative and thus simultaneously diagonalizable by a gauge transformation, so that we may write

\begin{equation}
\Phi^{m} = U\left( \begin{array}{ccccc} x_{1}^{m}  &          & & &0  \\
                                                   &x_{2}^{m} & & &   \\
                                                   &          &. & &    \\
                                                   &          & &. &     \\
                                           0       &          & & &x_{N}^{m} \end{array} \right) U^{-1},  \label{appAeqn13}
\end{equation}

 where the $N \times N$ unitary matrix ${\it U}$ is independent of m. 

The simultaneous, real eigenvalues $x_{i}^{m}$ give the positions of the N distinct D-branes in the m-th transverse direction. 
and the masses of the fields corresponding to the off-diagonal matrix elements are given precisely by the distances $|x_{i}-x_{j}|$ between the corresponding branes.

This description means that an interpretation of the D-brane configuration in terms of classical geometry is only possible in the classical ground state of the system, whereby the matrices become commutative.

One can deduce that at lower energies (larger distances) space-time is commutative and the matrices  $\Phi^{m}$ are simultaneously diagonalizable, giving the positions of the individual D-branes from their spectrum of eigenvalues, whilst at higher energies (short distances) space-time is non-commutative and the positions of the D-brane cannot be well defined, a result of the uncertainty principle.

\end{document}